\documentclass[twocolumn,floats,floatfix,amssymb,amsmath,nofootinbib,prd,superscriptaddress]{revtex4-2}
\usepackage{graphicx}
\usepackage{dcolumn}
\usepackage{bm}
\usepackage{amsmath,amssymb,bm,booktabs,color}
\usepackage{multirow}
\usepackage{lineno}
\usepackage[T1]{fontenc} 
\usepackage[utf8]{inputenc}
\usepackage{graphicx}
\usepackage{float}
\usepackage{multirow}
\usepackage{verbatim}
\usepackage{amsmath}
\usepackage{graphicx,subfigure,epsfig}
\usepackage{mathrsfs}
\usepackage{hyperref}
\hypersetup{colorlinks=true,citecolor=blue,linkcolor=blue,urlcolor=black,filecolor=blue}
\usepackage{color}
\usepackage{xcolor}

\begin{document}

\title{Black holes surrounded by dark matter spike: Spacetime metrics and gravitational wave ringdown waveforms}

\author{Dong Liu\href{https://orcid.org/0000-0001-7514-4592}{\includegraphics[width=0.3cm]{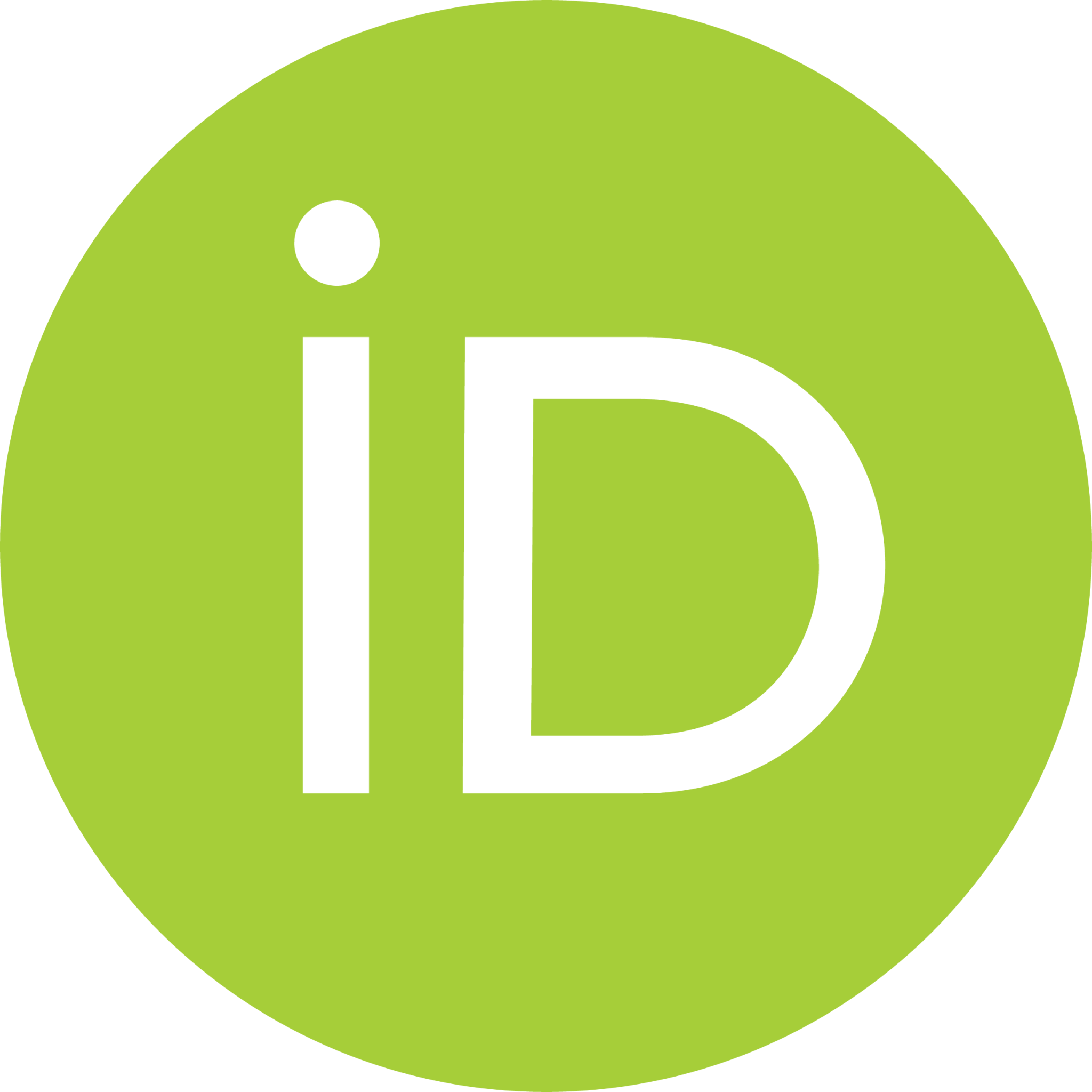}}}
\email{dongliu@gzmu.edu.cn}
\affiliation{Department of Physics, Guizhou Minzu University, Guiyang, 550025, China.}

\author{Yi Yang\href{https://orcid.org/0000-0003-1886-8716}{\includegraphics[width=0.3cm]{orcid.png}}}
\email{yiyang@mail.gufe.edu.cn}
\affiliation{School of Mathematics and Statistics, Guizhou University of Finance and Economics, Guiyang, 550025, China}

\author{Zheng-Wen Long\href{https://orcid.org/0000-0001-7945-2867}{\includegraphics[width=0.3cm]{orcid.png}}}
\email{zwlong@gzu.edu.cn}
\affiliation{College of Physics, Guizhou University, Guiyang, 550025, China}

\date{\today}

\begin{abstract}
The supermassive black holes at the centers of galaxies may be surrounded by dark matter spike, which could leave detectable imprints on the gravitational wave signals they emit. In this work, combining the mass model of M87, we present black hole analytical solutions in dark matter spike from the Tolman-Oppenheimer-Volkof equations. Meanwhile, the gravitational waves emitted at the ringdown phase of the black holes under the axial gravitational perturbation is investigated, and compare them with the Schwarzschild black hole. The main research methods are time domain integration method and continued fraction method. Besides, the impact of the different dark matter parameters on the quasinormal frequencies is investigated and its detectability on these impacts based on the space-based detectors. Our results indicate that the impacts of dark matter spike on the quasinormal frequencies of black holes can reach up to the order of $10^{-4}$. These results may provide some help in further exploring the impact of dark matter spike on the black holes and their related gravitational wave phenomena.

\end{abstract}

\maketitle

\section{Introduction}\label{s1}
The true nature of dark matter remains one of the most pressing and fundamental questions in modern physics, astronomy, and cosmology. Now, there is a large number of the observational evidence which strongly supports the existence of dark matter, such as galaxy rotation curves, the large-scale structure of the universe, and the cosmic microwave background radiation. Furthermore, the newest data from the Planck Results \cite{Planck:2018vyg} indicate that dark matter constitutes approximately 84.5\% of the total matter content of the universe and nearly 26.8\% of its total energy density, underscoring its critical role in cosmic evolution. Despite this abundance of indirect evidence, the particle properties of dark matter remains a profound mystery. To date, scientists have determined that dark matter does not interact via electromagnetic radiation but interacts gravitationally with ordinary matter \cite{Freese:2008cz,Navarro:1995iw,Clowe:2006eq}. This lack of electromagnetic interaction makes direct detection extraordinarily challenging, yet it also may provide a crucial pathway for probing its elusive properties.

On the other hand, the rapid advancement of gravitational wave astronomy has provided the new possibilities for exploring the mysterious matter in our universe \cite{Barack:2018yly,Cardoso:2019rvt, Bar:2019pnz}. In 2015, the LIGO/Virgo collaboration \cite{LIGOScientific:2016aoc} achieved the first direct detection of a gravitational wave event, originating from the merger of two nearly equal-mass black holes. This breakthrough observation marked the dawn of a new era in gravitational wave astronomy. Recently, the Event Horizon Telescope (EHT) collaboration \cite{EventHorizonTelescope:2019dse,EventHorizonTelescope:2022apq} released images of black holes at the centers of galaxies, offering direct observational evidence for their existence. Actually, the black holes at galactic centers do not exist in isolation. Instead, they are often surrounded by complex matter distributions, including galactic nuclei, accretion disks, dark matter, and even other planets \cite{Konoplya:2011qq}. These complex environments could leave detectable imprints on the gravitational waves signals emitted by black holes \cite{Yunes:2011ws,Cardoso:2020iji,Zwick:2022dih}. If dark matter indeed resides in the centers of galaxies, it may significantly influence the dynamics of the black holes and the propagation of gravitational waves. Meanwhile, some studies have been successively proposed exploring the possibility of using gravitational-wave experiments to detect dark matter \cite{Eda:2013gg,Macedo:2013qea,Traykova:2021dua,Li:2021pxf,Dai:2023cft}. Therefore, quantitatively studying the distribution of dark matter around the black holes and its impact on the generation and propagation of gravitational waves has emerged as an exciting and profound research direction. This endeavor not only promises to unveil the nature of dark matter but also holds the potential to provide critical insights into the formation of black holes and gravitational waves in our universe. 

To investigate how dark matter affects the black holes, it is essential to first understand the matter distribution around the black holes and their corresponding spacetime metrics. Z. Xu et al. \cite{Xu:2018wow,Xu:2021dkv} approached this problem from a Newtonian theory perspective, deriving the spacetime metrics of the black holes using the mass function. In contrast, Cardoso et al. \cite{Cardoso:2021wlq} solved the Einstein field equations to obtain the spacetime metrics of the black holes at galactic centers. Similarly, Jusufi \cite{Jusufi:2022jxu} proposed an exact solution analogous to the state equation for galactic matter, offering a novel thought for analyzing the dark matter distribution in galactic centers. These studies show that evaluating the impact of dark matter on gravitational wave propagation requires precise knowledge of dark matter profiles and accurate galactic parameters, such as those characterizing the M87 and the Milky Way. Gondolo et al. \cite{Gondolo:1999ef} showed that if dark matter exists in galactic centers, it could be accreted by the strong gravity of the central black hole, forming a density-enhanced spike structure. Subsequently, Sadeghian et al. \cite{Sadeghian:2013laa} incorporated relativistic modifications into this model, finding that the density of the dark matter spike could increase by 15\%. This further suggests that these structures could have the potential observational implications. Furthermore, other classic dark matter profiles, such as the NFW profile \cite{Navarro:1995iw,Liu:2021xfb}, Bondi profile \cite{Bondi:1952ni,Zhao:2023itk}, Pseudo-isothermal profile \cite{Begeman:1991iy,Yang:2023tip} and Einasto profile \cite{1965On, Wang:2019ftp}, have been widely used to describe the distribution of dark matter in the universe. These models may provide a robust theoretical foundation for further exploring the influence of dark matter on black holes and their associated gravitational wave phenomena.

From the time-domain signals of gravitational wave waveforms, the merger of binary black holes generally undergoes three stages, that is the inspiral phase, the merger phase, and the ringdown phase \cite{LIGOScientific:2016aoc}. Among these, the ringdown phase is characterized by oscillatory behavior with exponential decay, which can be described by a wave equation. During this phase, the quasinormal modes of the black holes dominate these signals, and these signals are entirely determined by the parameters of the black holes themselves. Therefore, the quasinormal modes are often referred as the characteristic sound of the black holes, serving as a vital tool for identifying and studying the black holes in our universe. The quasinormal modes of the black holes surrounded by dark matter were widely investigated in Refs. \cite{Zhang:2021bdr,Zhao:2023tyo,Chakraborty:2024gcr,Toshmatov:2025rln}. As a specific research case, here we focus on the method for solving the analytical solution of the black hole metric surrounded by dark matter spike and explore how varying parameters of the dark matter spike affect the quasinormal modes of the black holes, particularly in the context of the M87 galaxy. Additionally, we also analyze the potential of space-based detectors to observe these impacts. In this work, these analytical solutions for the black hole metric can be obtained by fixing specific parameter of dark matter spike. Through fixing these parameter, we can gain valuable insights into the interactions and dynamics involved between the dark matter and the black hole. These results may provide some help for investigating the impact of dark matter spike on the quasinormal modes of the black holes.

The motivation of this work is driven by three key theoretical and observational considerations. Firstly, we aim to enhance the self-consistency of the gravitational background. Unlike the previous work that treat the dark matter spike as a pure matter environment \cite{Liu:2024xcd}, we are interested in constructing a black hole metric strictly within the framework of full General Relativity. With the help of the Tolman-Oppenheimer-Volkoff equations, we seek to present the analytical solutions of black hole in dark matter spike and theirs impacts on the black hole spacetime. Secondly, we are motivated to investigate the potential impact of computational precision on physical conclusions. We note that the subtle modulation effects caused by DM spikes may be easily obscured by the intrinsic approximation errors of lower-order methods such as WKB or Prony methods. Therefore, we implement the high-precision continued fraction method to determine whether a more refined numerical approach can reveal subtle physical features that were previously underestimated or overlooked. Thirdly, under the framework of General Relativity, gravitational perturbations are typically triggered by the injection of gravitational waves or the capture of particles by a black hole. While the construction of black hole ringdown waveforms based on the space-based gravitational wave detectors necessitates the solving of the gravitational perturbation. Therefore, we are interested in determining whether the impacts of dark matter spike on black hole possesses sufficient detectability. Through these investigations, we aim to provide a robust theoretical framework and a more promising observational window for capturing the influence of dark matter spikes on black holes.

This work is organized as follows: In Sect. \ref{s2}, we introduce the dark matter distribution around the supermassive black hole and how to obtain the analytical solutions of black holes in dark matter spike from the TOV equations. In Sect. \ref{s3}, we present the wave equation of the black holes in dark matter spike under the axial gravitational perturbation. The numerical methods for calculating the quasinormal modes are given, i.e. the time domain integration method, Prony method and continued fraction method. In Sect. \ref{s4}, we investigate the quasinormal modes of the black holes in dark matter spike at the galactic center of M87, and then compare them with the Schwarzschild black hole. Besides, we examine the impacts of the different dark matter parameters on the quasinormal mode of the black holes. Finally, the detectability for the space-based detectors to detect these impacts of dark matter spike on the quasinormal modes is explored. The Sect. \ref{s5} is our summary. In this work, we always use the natural unit of $G=c=2 M_\text{BH}=1$, and the radius of the Schwarzschild black hole is $R_\text{s}=2 GM_\text{BH}/c^2$.

\section{The metric of black holes in dark matter spike}\label{s2}
Schwarzschild \cite{Schwarzschild:1916uq} was the first to solve Einstein field equations, providing an exact solution for a spherically symmetric vacuum, known as the Schwarzschild black hole. However, in the real universe, the vicinity of a black hole is rarely an ideal vacuum and is instead surrounded by complex matter distributions, such as galactic nuclei, dark matter, and other nearby planets. Therefore, in this section, we consider a spherically symmetric black hole surrounded by dark matter spike and then derive the corresponding modified black hole metric to better reflect the physical environment of the universe. 

Firstly, we assume a stress tensor of an anisotropic fluid in dark matter (DM) spike
\begin{equation}
T_{\nu }^{\mu}=\text{diag}(-\rho(r),P_r(r),P_t(r),P_t(r)),
\label{e1}
\end{equation}
where $\rho(r)$ is DM density distribution, $P_t(r)$ denotes tangential pressure and $P_r(r)=0$ is equivalent to assuming an anisotropic fluid in DM spike with vanishing radial pressure. The validity of this assumption in the context of DM spike was demonstrated by Daghigh et al. in Refs. \cite{Daghigh:2023ixh,Daghigh:2022pcr}. In 4D spherically symmetric spacetime, the metric of a black hole can generally be expressed as
\begin{equation}
\begin{aligned}
ds^{2}=-f(r)dt^{2}+ \frac{dr^{2}}{1-2m(r)/r}+r^2d\Omega ^2,
\label{e2}
\end{aligned}
\end{equation}
where $d\Omega ^2=d\theta ^{2}+\sin^{2}\theta d\phi ^{2}$, the radial function $f(r)$ and the mass function $m(r)$ can be obtained from the Tolman-Oppenheimer-Volkoff (TOV) equations
\begin{equation}
\frac{dm(r)}{dr}=4\pi r^2\rho(r),
\label{e3}
\end{equation}
and
\begin{equation}
\frac{f'(r)}{f(r)}=\frac{2m(r)}{r^2-2rm(r)},
\label{e4}
\end{equation}
while the tangential pressure $P_t(r)$ in the TOV equations reads
\begin{equation}
\begin{aligned}
2P_t(r)=-\frac{\rho(r)m(r)}{r-2m(r)}.
\label{e5}
\end{aligned}
\end{equation}
From Eqs.(\ref{e3})-(\ref{e5}), we find that in solving for non-vacuum situations, the specific density distribution plays a crucial role in solving the TOV equations. Once the density distribution $\rho(r)$ is determined, the matter properties (e.g., $m(r),f(r),P(r)=P_t(r)$) around the black hole will become clear. In the next subsections, we mainly discuss the density distribution of dark matter around the central black hole and solve the TOV equation for the black hole solutions.

\begin{figure*}[t!]
\centering
{
\includegraphics[width=.95\columnwidth]{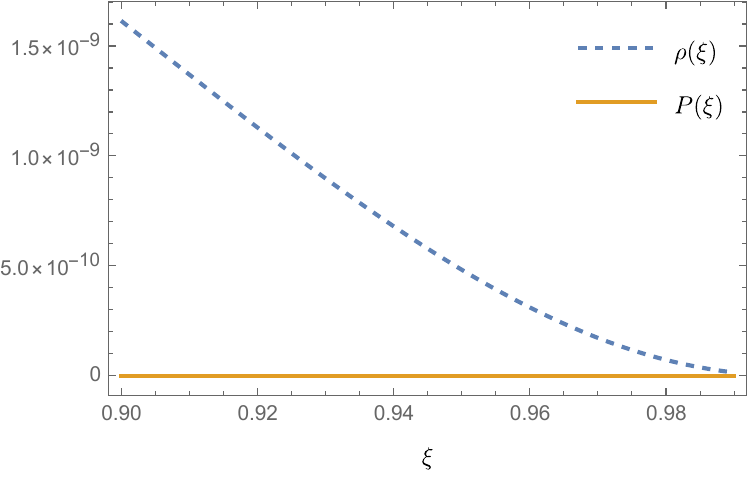}
}
{
\includegraphics[width=.9\columnwidth]{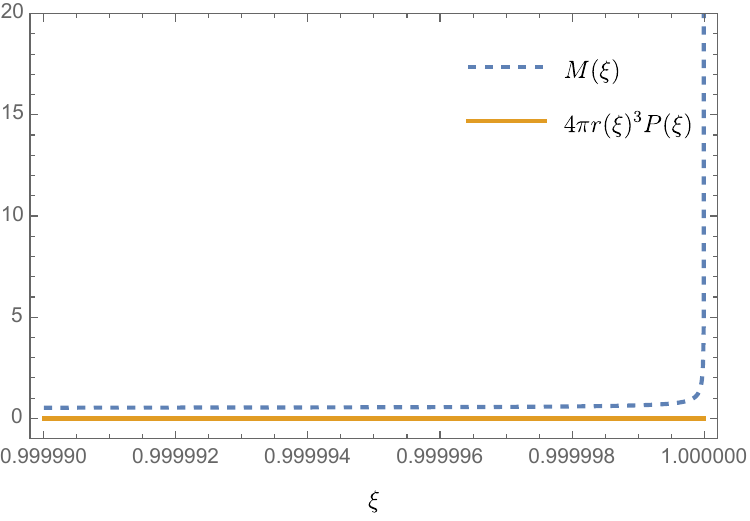}
}
\caption{The DM spike pressure $P(\xi)$ and density $\rho(\xi)$ are given as functions of $\xi=1-2M_\text{BH}/r$ in the fully relativistic case (left panel), where all variables are expressed using black hole parameters. Besides, the functions $M(\xi)$ and $4\pi r(\xi)^3P(\xi)$ are plotted to demonstrate that the pressure term remains negligible as \(\xi\) approaches 1 (\(r \to \infty\)) and \(M(\xi)\) increases from \(M_{\rm BH}\) to \(M_{\rm BH}\)+\(M_{\rm sp}\)(max) (right panel). The parameters we used are $M_\text{BH}=1/2$, $\alpha_\gamma=0.1$, $\rho_0=6.9 \times 10^6 M_\odot/\text{kpc}^3$, $r_0=91.2 \text{kpc}$.}
\label{nf1}
\end{figure*}

\subsection{The dark matter distribution around the supermassive black hole}
Dark matter typically forms a density distribution called spike structure because of the strong gravity of the supermassive black hole in the galaxy's center \cite{Gondolo:1999ef,Sadeghian:2013laa}
\begin{equation}
\rho_\text{spike}(r)=\rho_\text{sp}\left(1- r_\text{b} / r\right)^{3}\left( R_{\text{sp}} / r\right)^{\gamma _\text{sp} },
\label{e6}
\end{equation}
where $\gamma _\text{sp}=(9-2\gamma )/(4-\gamma )$, $r_\text{b}$ is inner boundary of the dark matter spike ($r_\text{b}=2R_\text{s}$ and $r_\text{b}=4R_\text{s}$ correspond to the fully relativistic case \cite{Sadeghian:2013laa,Tang:2020jhx} and the Newtonian approximation \cite{Gondolo:1999ef}, while $R_\text{s}$ is Schwarzschild radius). To be specific, for the spherically symmetric black holes, the dark matter density reaches its peak near $r\geq r_\text{b}$. But it has a steep cutoff at $r= r_\text{b}$ and below this value, the dark matter density vanishes because of annihilation or dropping into the black holes. The symbols $\rho_\text{sp}$ and $R_\text{sp}$ represent the density and the radius of the dark matter spike, respectively, as given below:
\begin{equation}
\begin{aligned}
R_\text{sp} = \alpha _\gamma r_0 \left( \frac{M_{\text{BH}}}{\rho_0r_0^3} \right)^{\frac{1}{3-\gamma }}, \quad \rho_ \text{sp} = \rho_0 \left( \frac{R_\text{sp}}{r_0} \right)^{-\gamma },
\label{e7}
\end{aligned}
\end{equation}
where $\gamma$ is the power-law index, $\alpha_\gamma$ is normalization scale factor, $M_\text{BH}$ is the mass of the central black hole, and the symbols $\rho_0$, $r_0$ represent density and radius of dark matter halo in the galaxy's center, respectively.

Based on Eq. (\ref{e6}), the mass distribution of the dark matter spike can be expressed as
\begin{equation}
\begin{aligned}
M_{\text{sp}}(r)=4 \pi \int_{r_\text{b}} ^r    r'^2\rho_\text{spike}(r')dr',  \quad r_\text{b} \leq r' \leq R_{\text{sp}},
\label{e8}
\end{aligned}
\end{equation}
and then the total mass of black hole and dark matter is
\begin{equation}
m(r)=\begin{cases}
\quad \quad \text{ $M_\text{BH},$ } \quad \quad  \quad \quad  r \leq  r_\text{b}, \\ 
\text{ $M_\text{BH}+M_\text{sp}(r),$ } \quad r_\text{b} \leq r \leq R_\text{sp},\\ 
\text{ $M_\text{BH}+M_\text{sp},$ } \quad \quad \quad r > R_\text{sp}.
\end{cases}
\end{equation}
Therefore, based on the Eq. (\ref{e8}), the mass of dark matter spike becomes
\begin{equation}
\begin{aligned}
&M_\text{sp}(r) = 4 \pi \rho_\text{sp} \left( 
\frac{6 r_\text{b}^3 \left(R_\text{sp}/r_\text{b}\right)^{\gamma_\text{sp}}}{\gamma_\text{sp}(\gamma_\text{sp}-1)(\gamma_\text{sp}-2)(\gamma_\text{sp}-3)} +
\right.\\
&\left. 
\left( \frac{r^3}{3-\gamma_\text{sp}} + \frac{3 r^2 r_\text{b}}{\gamma_\text{sp}-2} - \frac{3 r r_\text{b}^2}{\gamma_\text{sp}-1} + \frac{r_\text{b}^3}{\gamma_\text{sp}} \right) 
\left( \frac{R_\text{sp}}{r} \right)^{\gamma_\text{sp}} 
\right).
\label{e10}
\end{aligned}
\end{equation}
In the following discussions, we mainly utilize the mass model of M87 for detailed research and analysis, focusing on its implications for the surrounding dark matter distribution and black hole dynamics. At the galactic center of M87, the mass of the central black hole is $M=6.5 \times 10^9 M_\odot$ \cite{EventHorizonTelescope:2019dse}. The density of the dark matter halo is $\rho_0=6.9 \times 10^6 M_\odot/\text{kpc}^3$, and the characteristic radius is $r_0=91.2 \text{kpc}$ \cite{Jusufi:2019nrn,Xu:2021dkv}. Here, we can use Schwarzschild black holes as the black hole units. Under black hole units (BHU), the density and characteristic radius of dark matter halo can be converted by the following formula:
\begin{equation}
r_\text{0}(\text{BHU})=r_\text{0} / (2 G M_{\text{BH}}/c^2) \times r_\text{BH},
\label{e11}
\end{equation}
and
\begin{equation}
\rho_\text{0}(\text{BHU})=\rho_\text{0} /(M_\text{BH}/(4/3 \pi (2 G M_{\text{BH}}/c^2)^3))\times \rho_{\text{BH}}.
\label{e12}
\end{equation}
After the conversions, we have the following parameters in the black hole units, $\rho_0 (\text{BHU})  \approx 1.41\times10^{-22}$, $r_0 (\text{BHU}) \approx 1.42 \times 10^8$. Besides, we adopt the power-law index $\gamma$ from Refs. \cite{McMillan:2016jtx, Shen:2023kkm,Nampalliwar:2021tyz} and mainly focus on the case of $\gamma =  [0,1/2,1]$ and then $\gamma_\text{sp} = [9/4,16/7,7/3]$. The selection of these values is based on numerical simulations of collisionless dark matter particles in galactic halos. While the normalization $\alpha_\gamma$ is a small value and here we choose $\alpha_\gamma=1/10$. With these values, the impacts of dark matter spike on the black holes can be quantitatively analyzed, providing deeper insights into the interaction between dark matter density profiles and black hole properties. Here, to ensure that such an analysis is grounded in a physical framework, it is imperative to first examine the rationality of the underlying gravitational model. To this end, we firstly present the evolution of the DM spike pressure $P(\xi)$ and density $\rho(\xi)$ as a function of $\xi = 1 - 2M_{\text{BH}}/r$ in Fig. \ref{nf1}. The left panel clearly demonstrates that the DM density $\rho(\xi)$ remains several orders of magnitude higher than the pressure $P(\xi)$ ($P(\xi) \ll \rho(\xi)$) throughout the region of interest. This confirms that the pressure effect of the DM spike may be negligible in this astrophysical environment.In the right panel of Fig. \ref{nf1}, we compare the mass function $M(\xi)$ with the pressure-related term $4\pi r(\xi)^3 P(\xi)$. As $\xi$ approaches 1 (corresponding to $r \to \infty$), $M(\xi)$ increases monotonically from the black hole mass $M_{\text{BH}}$ to $M_{\text{BH}} + M_{\text{sp}}(\text{max})$, while the pressure term remains consistently near zero. These numerical results provide strong justification for the vanishing pressure approximation adopted in our metric derivation. Such self-consistency indicates that the gravitational influence of the DM spike is dominated by its mass profile, while the pressure's contribution to the spacetime curvature remains negligible within the chosen parameter space.

\subsection{The black hole solutions of the TOV equations in dark matter spike}
\begin{figure}[t!]
\centering
{
\includegraphics[width=.95\columnwidth]{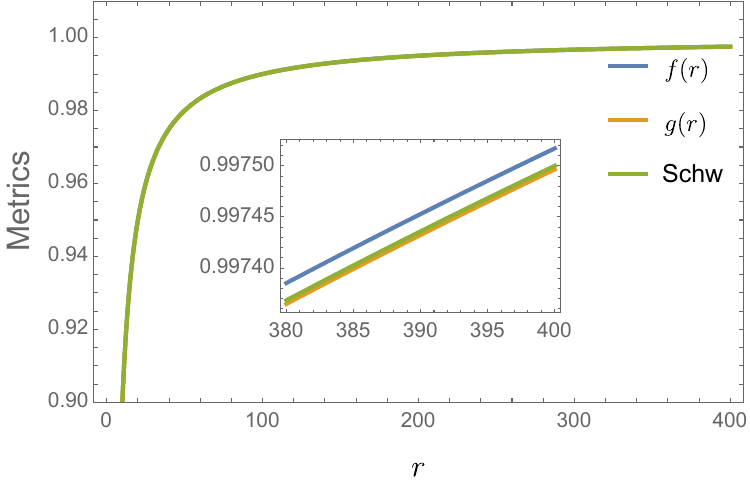}
}
\caption{The black hole metrics as a function of the radial coordinate $r$ under the fully relativity. The parameters we used are $M_\text{BH}=1/2$, $\gamma=1$, $r_\text{b}=2$, $\alpha_\gamma=0.1$, $\rho_0=6.9 \times 10^6 M_\odot/\text{kpc}^3$, $r_0=91.2 \text{kpc}$.}
\label{f1}
\end{figure}

\begin{figure}[t!]
\centering
{
\includegraphics[width=.95\columnwidth]{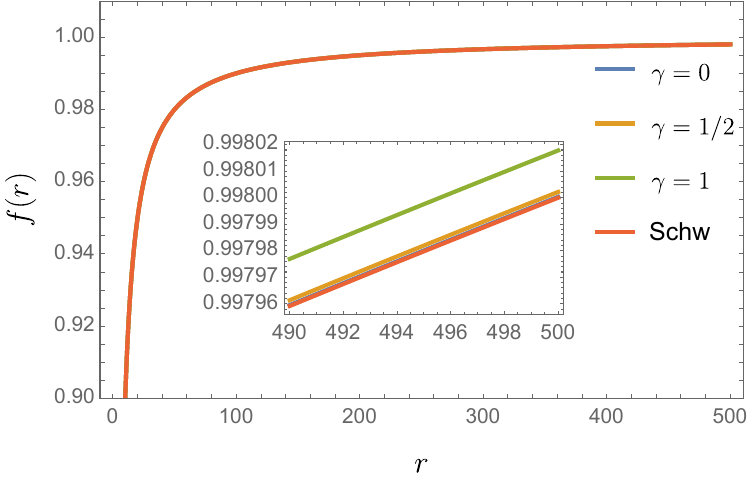}
}
{
\includegraphics[width=.95\columnwidth]{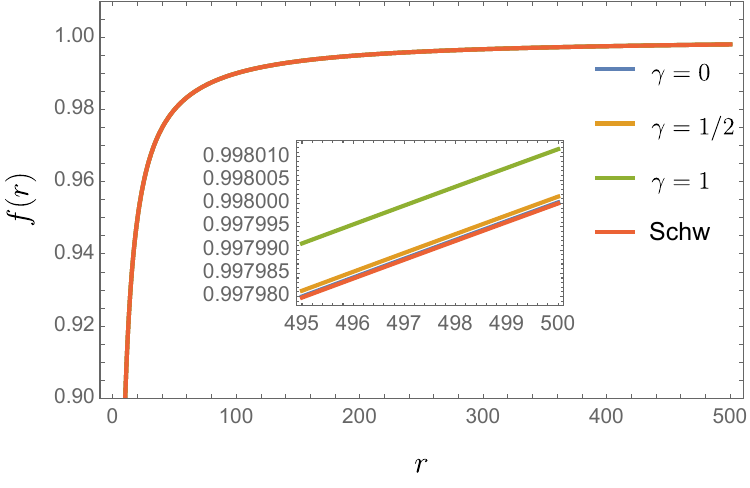}
}
\caption{The black hole metrics for the different power-law index $\gamma$ as a function of the radial coordinate $r$ under the the fully relativity (top panel) and the Newtonian approximation (bottom panel). The parameters we used are $M_\text{BH}=1/2$, $\alpha_\gamma=0.1$, $\rho_0=6.9 \times 10^6 M_\odot/\text{kpc}^3$, $r_0=91.2 \text{kpc}$.}
\label{f2}
\end{figure}

In this subsection, we give the black hole metric in dark matter spike by solving the TOV equation. Firstly, we write metric as
\begin{equation}
\begin{aligned}
ds^{2}=-f(r)dt^{2}+ g(r)^{-1}dr^{2}+r^2d\Omega ^2,
\label{e13}
\end{aligned}
\end{equation}
and
\begin{equation}
g(r)=1-\frac{2m(r)}{r}=1-\frac{2M_\text{BH}}{r}-\frac{2M_\text{sp}(r)}{r},
\label{e14}
\end{equation}
where $M_\text{sp}(r)$ can be obtained in Eq. (\ref{e10}). Now, the function $f(r)$ in the black hole metric in dark matter spike remains unknown, but it can be determined from TOV equations. The simplest method to obtain the metric function $f(r)$ is directly through integral Eq. (\ref{e4}), but we find it fails after our attempt. However, with the help of Eq. (\ref{e14}), one can rewrite Eq. (\ref{e4}) as
\begin{equation}
\begin{aligned}
\frac{f'(r)}{f(r)}=-\frac{1}{r}+\frac{1}{r-2(M_\text{BH}+A(r))},
\label{e15}
\end{aligned}
\end{equation}
where, the function $A(r)$ and its constant coefficients $a_1$, $a_2$, $a_3$, $a_4$, $a_5$ are as follows:
\begin{equation}
\begin{aligned}
&A(r)=a_1r^{3-\gamma _{\text{sp}}}+a_2r^{2-\gamma _{\text{sp}}}+a_3r^{1-\gamma _{\text{sp}}}+a_4r^{-\gamma _{\text{sp}}}+a_5,\\
&a_1=-\frac{4 \pi \rho_{\text{sp}} R_{\text{sp}}^{\gamma_{\text{sp}}}}{\gamma_{\text{sp}}-3}, \quad a_2=\frac{12 \pi \rho_{\text{sp}} r_\text{b} R_{\text{sp}}^{\gamma_{\text{sp}}}}{\gamma_{\text{sp}}-2}, \\
&a_3=-\frac{12 \pi \rho_{\text{sp}} r_\text{b}^2 R_{\text{sp}}^{\gamma_{\text{sp}}}}{\gamma_{\text{sp}}-1}, \quad a_4=\frac{4 \pi \rho_{\text{sp}} r_\text{b}^3 R_{\text{sp}}^{\gamma_{\text{sp}}}}{\gamma_{\text{sp}}},\\
&a_5=\frac{24 \pi \rho_{\text{sp}} r_\text{b}^3 (R_{\text{sp}}/r_\text{b})^{\gamma_{\text{sp}}}}{\gamma_{\text{sp}}(\gamma_{\text{sp}}-1)(\gamma_{\text{sp}}-2)(\gamma_{\text{sp}}-3)}.
\label{e16}
\end{aligned}
\end{equation}
Now, if the power-law index $\gamma$ is fixed, with $\gamma=1 (\gamma_{\text{sp}}=7/3)$, Eq. (\ref{e15}) can be integrated, we obtain
\begin{equation}
\ln [f(r)]=\int -\frac{1}{r}+\int \frac{1}{r-2(M_\text{BH}+A(r))}+C,
\label{e17}
\end{equation}
after integration, the final analytic expression for the black hole metric, $f(r)=\exp[\ln [f(r)]]$, is
\begin{equation}
f(r)=(\frac{r_\text{b}-2M_\text{BH}}{r})\times \exp \left[\frac{3}{2}\times [\mathcal{F}(r)-\mathcal{F}(r_\text{b})] \right],
\label{e18}
\end{equation}
where, $\mathcal{F}(r)$ and $\mathcal{F}(r_\text{b})$ are
{\small
\begin{equation}
\begin{aligned}
&\mathcal{F}(r)= \sum_{x_i} \frac{ \log[(r^{1/3}-x_i)]\times  x_i^7}{-3a_3-6a_2x_i^3-7a_5x_i^4-7M_\text{BH}x_i^4-9a_1x_i^6+5x_i^7},\\
&\mathcal{F}(r_\text{b})= \sum_{x_i} \frac{ \log[(r_\text{b}^{1/3}-x_i)]\times  x_i^7}{-3a_3-6a_2x_i^3-7a_5x_i^4-7M_\text{BH}x_i^4-9a_1x_i^6+5x_i^7},
\label{e19}
\end{aligned}
\end{equation}}
while $x_i$ is the root of the following equation
\begin{equation}
\begin{aligned}
2a_4+2a_3x^3&+2a_2x^6+2a_5x^7\\
&+2M_\text{BH}x^7+2a_1x^9-x^{10}=0.
\label{e20}
\end{aligned}
\end{equation}
while the integration constant $C$ can be determined by the initial condition $f(r_\text{b})=1-2M_\text{BH}/r_\text{b}$. In Fig. \ref{f1}, we first present this black hole metrics as a function of $r$ and compare it to the Schwarzschild black hole. Similarly, when power-law index $\gamma$ is assigned other values, the black hole metric $f(r)$ can be obtained through integration. For simplicity, in Appendix \ref{A}, we present a summary of the black hole metrics $g(r)$ and $f(r)$ for the different power-law indexes $\gamma=0$, $\gamma=1/2$ and $\gamma=1$. Meanwhile, in Fig. \ref{f2}, we also present the function images of the black hole metrics $f(r)$ under the different power-law indexes. For comparison, the case of the Schwarzschild black hole is represented with an orange line.

\section{Axial gravitational perturbation of black holes in dark matter spike}\label{s3}
\subsection{The wave equation for the axial gravitational perturbation}
Currently, the perturbations of the black holes can be excited by different spin fields, such as scalar field perturbation, electromagnetic field perturbation and gravitational perturbation. Here, we primarily focus on axial gravitational perturbation, as it can describe the perturbation caused by the emission of gravitational waves or the infall of particles into the black hole \cite{Davis:1972ud,1988sfbh.book.....F}. Therefore, in gravitational perturbation, the perturbed metric can be described as the sum of the background metric $\bar{g}_{\mu \nu }$ and a small linear perturbation $h_{\mu \nu }$, 
\begin{eqnarray}
g_{\mu \nu }=\bar{g}_{\mu \nu }+h_{\mu \nu },
\label{e21}
\end{eqnarray}
where, $\bar{g}_{\mu \nu }$ is a black hole metric in a dark matter spike in Eq. (\ref{e13}) and $h_{\mu \nu }$ reads
\begin{equation}
h_{\mu\nu}=\sum_{l=0}^{\infty }\sum_{m=-l}^{m=l}[(h_{\mu\nu}^{lm})^{\text{axial}}+(h_{\mu\nu}^{lm})^{\text{polar}}].
\label{e22}
\end{equation}
Here, we primarily focus on the axial gravitational perturbation and this small linear perturbation $(h_{\mu\nu}^{lm})^{\text{axial}}$ in Ref. \cite{Regge:1957td} can be defined as
\begin{eqnarray}
(h_{\mu\nu}^{lm})^{\text{axial}}=\begin{pmatrix}
0 &  0& 0 & h_{0}\\
 0& 0 &  0& h_{1}\\
0 & 0 & 0 &0 \\
h_{0} &h_{1}  &0  & 0
\end{pmatrix}{\rm sin}\theta \partial \theta P_{l}({\rm cos}\theta ),
\label{e23}
\end{eqnarray}
where, $h_0$, $h_1$ denote the functions  $h_0(t,r)$, $h_1(t,r)$, $P_{l}({\rm cos}\theta )$ is the Legendre polynomials of the order $l$ and $m=0$ is the case for the spherical symmetry. Here, the wave equation of the axial gravitational perturbation can be obtained from Einstein field equations
\begin{equation}
\begin{aligned}
E_{\mu\nu} \equiv R_{\mu\nu}-\frac{1}{2} g_{\mu\nu}R=\kappa T_{\mu\nu},
\label{e24}
\end{aligned}
\end{equation}
where, $R_{\mu \nu }$ is Ricci tensor, $g_{\mu\nu}$ is the perturbed metric, $R$ is Ricci scalar and $T_{\mu\nu}$ is the energy-momentum tensor of dark matter spike. Based on insights from Refs. \cite{Zhang:2021bdr,Zhao:2023tyo}, and neglecting the perturbation of dark matter, the nonzero components $E_{24}$ and $E_{34}$ can be expressed as
\begin{equation}
\begin{aligned}
\frac{\partial^2}{\partial t ^2}h_1 &-\frac{\partial}{\partial r}\frac{\partial}{\partial t}h_0 +\frac{2}{r} \frac{\partial}{\partial t}h_0-\frac{2g(r)f'(r)}{2r}h_1 \\
& +\frac{ f(r) [2 l (l+1) - 2rg'(r) - 4 g(r) ]}{2 r^2}h_1=0,
\label{e25}
\end{aligned}
\end{equation}
and
\begin{equation}
\begin{aligned}
-\frac{1}{f(r)}\frac{\partial }{\partial t} h_0 +  g(r) \frac{\partial}{\partial r}h_1 + \frac{[f(r)\times   g(r)]' }{2 f(r)}h_1=0,
\label{e26}
\end{aligned}
\end{equation}
where, the prime symbols $'$ and $''$ are respectively the first and second derivative with respects to $r$. Solving the term $\frac{\partial }{\partial t} h_0$ in Eq. (\ref{e26}) and taking the result into Eq. (\ref{e25}), then using the following ansatz
\begin{equation}
\begin{aligned}
\psi (t,r)=\sqrt{f(r)g(r)}h_1(t,r)/r,
\label{e27}
\end{aligned}
\end{equation}
the Eqs. (\ref{e25}) and  (\ref{e26}) can be rewritten as a new form
\begin{equation}
\begin{aligned}
\frac{\partial^2 }{\partial t^2} \psi(t,r) &- fg\frac{\partial^2 }{\partial r^2} \psi(t,r) -\frac{gf'+fg'}{2}\frac{\partial }{\partial r} \psi(t,r) \\
&+ V(r) \psi(t,r)=0.
\label{e28}
\end{aligned}
\end{equation}
Now, we use tortoise coordinates to replace the radial coordinates in Eq. (\ref{e28}), because it can extend the event horizon to the infinity. Here, the tortoise coordinates can be described as
\begin{equation}
\begin{aligned}
dr_*=\frac{1}{\sqrt{f(r)g(r)}}dr.
\label{e29}
\end{aligned}
\end{equation}
Finally, the wave equation for the axial gravitational perturbation in dark matter spike can be written as
\begin{equation}
\begin{aligned}
\frac{\partial^2 }{\partial t^2} \psi(t,r) - \frac{\partial^2 }{\partial r_*^2} \psi(t,r)+ V (r) \psi(t,r)=0,
\label{e30}
\end{aligned}
\end{equation}
where, the effective potential $V(r)$ reads
\begin{equation}
\begin{aligned}
V (r)&=\frac{3f(r)g(r)}{r^2} - \frac{2r[f(r)\times g(r)]'\times 4f(r)g(r)}{4r^2} \\
&- \frac{2rg(r)f'(r) -f(r) [2 l (l+1)-2rg'(r) -4g(r) ]}{2 r^2}.
\label{e31}
\end{aligned}
\end{equation}
The functions $f(r), g(r)$ can be obtained in Eq. (\ref{e13}). In the Eqs. (\ref{e30}) and (\ref{e31}), one can find that if dark matter is absent ($\rho_\text{sp}=0$), the effective potential will degenerate into the case of Schwarzschild black hole, that is
\begin{equation}
\begin{aligned}
\left [ \frac{\partial^2 }{\partial t^2}  - \frac{\partial^2 }{\partial r_*^2}+ \left(1-\frac{2M}{r}\right)\left(\frac{l(l+1)}{r^2}-\frac{6M}{r^3}\right) \right ] \psi=0,       
\label{e32}
\end{aligned}
\end{equation}
where $r_*$ is giving by
\begin{equation}
dr_*=\frac{1}{1-2M/r}dr.
\label{e33}
\end{equation}

\subsection{Time domain integration method}\label{s5}
The behavior of the black hole perturbation in dark matter spike can be described by the wave equation in Eq. (\ref{e30}). While the quasinormal mode is a solution to this equation, which can be characterized by some complex frequencies. Before studying quasinormal frequencies, we first are interested in the time evolution of the quasinormal mode, as they provide valuable insights into the perturbative behavior near the black hole. Now, after introducing the light-cone coordinates $u=t-r_*$ and $v=t+r_*$ in Eq. (\ref{e30}), and we obtain 
\begin{equation}
4\frac{\partial ^{2}\psi (u ,v )}{\partial u \partial v } + V(u ,v)\psi (u ,v )=0,
\label{e34}
\end{equation}
where, $r_*$ is tortoise coordinates. For the Eq. (\ref{e34}), a reasonable discretization scheme in Refs. \cite {PhysRevD.63.084014,PhysRevD.64.044024,PhysRevD.72.044027} is described as
\begin{equation}
\begin{aligned}
\psi(N)&=\psi(W)+\psi(E)-\psi(S)-h ^2\frac{V(W)\psi(W)}{8}\\
&-h ^2\frac{V(E)\psi(E)}{8}+O(h ^4),
\label{e35}  
\end{aligned}                                           
\end{equation}
here, $h$ represents the step size of each grid cell, and the integration grid consists of $ N = (u + h, v + h)$, $W = (u + h, v)$, $E = (u, v + h)$, $S = (u, v)$, respectively. The Gaussian pulse is then applied to the two null surfaces, $u = u_0$ and $v = v_0$,
\begin{equation}
\begin{aligned}
\psi (u=u _{0} ,v )={\rm exp}\left [ -\frac{(v -v _{0})^{2}}{\sigma ^{2}} \right ],  \psi (u ,v=v_{0} )=0,  
\label{e36}     
\end{aligned} 
\end{equation}
where, we choose $v_0=10$ and $\sigma=3$. Thus, the time evolution of a black hole under axial gravitational perturbation can be determined. Finally, the quasinormal frequencies can be estimated from the time evolution using the Prony method \cite{RevModPhys.83.793,PhysRevD.75.124017,Chowdhury:2020rfj}, as shown in the following expression
\begin{eqnarray}
\psi(t)\simeq \sum ^p_{i=1}C_ie^{-i\omega_it}.                  
\label{e37}                    
\end{eqnarray}

\subsection{Continued fraction method}
Compared to the Prony method, the continued fraction method is currently the most accurate technique for calculating the quasinormal mode frequencies of the black holes. Since the impacts of dark matter on the quasinormal modes of black holes may be relatively small, precise numerical methods can provide more reliable results for the calculation of quasinormal modes. This method has been described by Leaver \cite{Leaver:1985ax,Leaver:1990zz} in detail, and further explored in other studies \cite{Berti:2009kk,Konoplya:2011qq,Anacleto:2021qoe,Siqueira:2022tbc}, all of which suggest it is a method with highly precision. Now, using the ansatz $\psi(t,r)=e^{- i \omega t}R(r)$, after rewriting Eq. (\ref{e30})  in terms of the usual radial coordinate, we obtain
\begin{equation}
\begin{aligned}
2r(r-r_\text{h})(r-r_\text{f}) \frac{\mathrm{d} ^2}{\mathrm{d} r^2}R(r)+r_\text{h}(r-r_\text{f})\frac{\mathrm{d} }{\mathrm{d} r}R(r)\\
+r_\text{f}(r-r_\text{h})\frac{\mathrm{d} }{\mathrm{d} r}R(r)+2r^3(\omega^2-V(r))R(r)=0,
\label{e38}
\end{aligned}
\end{equation}
where, $r_\text{h}$ is event horizon of the black hole in dark matter spike and $r_\text{f}$ are given by
\begin{equation}
f(r) \simeq  \frac{r-r_\text{f}}{r}, \quad g(r) \simeq  \frac{r-r_\text{h}}{r}.
\label{e39}
\end{equation}
where,  we choose $r_\text{f} \approx r_\text{h}$ in the numerical calculation, as the difference between them is approximately less than the order of $10^{-14}$, which can be negligible relative to the main contribution itself. Besides, based on the principle that nothing can escape from the black holes, the boundary conditions are simplified as the pure outgoing wave ($e^{i \omega r_*}$) at infinity and the pure incoming wave ($e^{-i \omega r_*}$) at the event horizon. While $r_*$ is the tortoise coordinates in the Eq. (\ref{e29}), and it can be described explicitly as $r_*=r+r_\text{h}\ln (r-r_\text{h})$. Therefore, near these two boundaries, the ringdown waveforms can be rewritten as
\begin{equation}
\begin{aligned}
r &\rightarrow \infty:\\
R(r)&=e^{i \omega r_*} \sim e^{i \omega (r+r_\text{h}\ln r)}\sim r^{i \omega r_\text{h}} e^{i \omega r},  \\
r &\rightarrow r_\text{h}:\\
R(r)&=e^{-i \omega r_*} \sim e^{-i \omega r_\text{h} \ln (r-r_\text{h})}\sim (r-r_\text{h})^{-i \omega r_\text{h}}, 
\end{aligned}
\label{e40}
\end{equation}
Taking Eq.(\ref{e40}) into account, this radial function $R(r)$ to Eq.(\ref{e38}) can be described as
\begin{equation}
R(r)=(r-r_\text{h})^{-i \omega r_\text{h}}r^{2i \omega r_\text{h}} e^{i \omega (r-r_\text{h})}\sum_{n=0}^{\infty}b_n\left ( \frac{r-r_\text{h}}{r} \right )^n,
\label{e41}
\end{equation}
Putting Eq.(\ref{e41}) into Eq.(\ref{e38}) for calculation, one can find that $b_n$ must satisfy the following three-term recurrence relation,
\begin{equation}
\begin{cases}
 & \alpha _0b_1+\beta _0b_0=0, \\
 &  \alpha _nb_{n+1}+\beta _nb_n+\gamma _nb_{n-1}=0, \quad n\geq 1
\end{cases}
\label{e42}
\end{equation}
where, we choose $b_0=1$ and the coefficients $\alpha _n$, $\beta _n$, $\gamma _n$ can be read as
\begin{equation}
\begin{aligned}
\alpha_n &= 1 + 2n + n^2 - 2i \omega r_\text{h} - 2n i \omega r_\text{h}, \\
\beta_n &= -1 - l(l+1) - 2n - 2n^2 + 4i \omega r_\text{h} \\
         & + \sigma ^2 + 8n i \omega r_\text{h} + 8 \omega^2 r_\text{h}^2, \\
\gamma_n &= n^2 - \sigma ^2  - 4n i \omega r_\text{h} - 4 \omega^2 r_\text{h}^2.
\end{aligned}
\label{e43}
\end{equation}
If we define the convergence series $S_{n} = b_{n}/b_{n-1}$, the recurrence relation can be rewritten as
\begin{equation}
S_n=\frac{-\gamma _n}{\beta _n+\alpha _nS_{n+1}}.
\label{e44}
\end{equation}
With Eq.(\ref{e42}), we also have $S_1=b_1/b_0=-\beta_0/\alpha_0$. So, we can get the main equation of the continued fraction method
 \begin{equation}
0=\beta _0-\frac{\alpha _0\gamma _1}{\beta _1-\frac{\alpha _1\gamma _2}{\beta _2-\frac{\alpha _2\gamma _3}{\beta _3-\cdots  }}}.
\label{e45}
\end{equation}
Given the value of the horizon $r_\text{h}$ and the angular quantum number $l$, the master equation of the continued fraction method is the equation for the quasinormal frequency $\omega$. In other words, $\omega$ is the root of a continued fraction equation or any of its inversions \cite{Leaver:1985ax,Berti:2005gp}. Then, we can obtain the value of the quasinormal frequency $\omega$ through the FindRoot algorithm, and the initial guess is the quasinormal frequency of the Schwarzschild black hole. 

Finally, the series $S_n$ needs to be truncated at $n\in N$ to ensure the convergence of this master equation \cite{Yoshino:2013ofa}. Concerning the convergence equation, Nollert proposed several solutions to ensure the convergence rate of this method \cite{Nollert:1993zz,Zhidenko:2006rs}. Therefore, from the recurrence relation (\ref{e42}) and the Eq. (\ref{e43}), one can find that the convergence series $b_{n+1}/b_n$ satisfies the relation at the order of $n^{-1}$,
\begin{equation}
\begin{aligned}
\frac{b_{n+1}}{b_n} = 1 + \frac{\sqrt{-2 i \omega r_\text{h}}}{\sqrt{n}}-\frac{3+8 i \omega r_\text{h}}{4n}.
\label{e46}
\end{aligned}
\end{equation}

\begin{table*}[t!]
\setlength{\abovecaptionskip}{0.2cm}
\setlength{\belowcaptionskip}{0.2cm}
\setlength{\tabcolsep}{12pt}  
\renewcommand{\arraystretch}{1.2}
\centering
\caption{The fundamental quasinormal modes of Schwarzschild black hole in the axial gravitational perturbation using the Prony method, continued fraction method and the Leaver's results in Ref. \cite{Leaver:1985ax,Iyer:1986nq}. The parameter we used is $M_\text{BH}=1/2$.}
\begin{ruledtabular}
\begin{tabular}{cccc}
      Modes $(M_{\text{BH}}=1/2) $                       &  Leaver's results              & Prony method                   & continued fraction method                  \\
 \hline
$l =2$                   & 0.747343 - 0.177925 i & 0.7487774752 - 0.1784811211 i & 0.7473433688 - 0.1779246314 i \\
$ l =3$                   & 1.198887 - 0.185406 i & 1.1916271155 - 0.1855550572 i & 1.1988865770 - 0.1854060959 i \\
$ l =4$                   & 1.6184 - 0.1884 i & 1.6118981130 - 0.1877619400 i & 1.6183567550 - 0.1883279220 i \\
\end{tabular}
\end{ruledtabular}
\label{t1}
\end{table*}

\begin{figure*}[t!]
\centering
\subfigure[Schwarzschild black hole]{\label{fig:a}
\includegraphics[width=.65 \columnwidth]{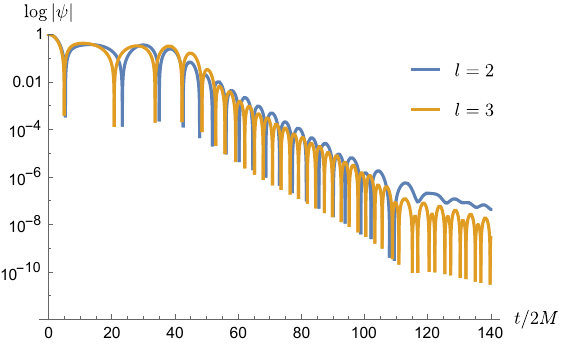}
}
\subfigure[Dark matter spike, $r_\text{b}=2$]{\label{fig:b}
\includegraphics[width=.65 \columnwidth]{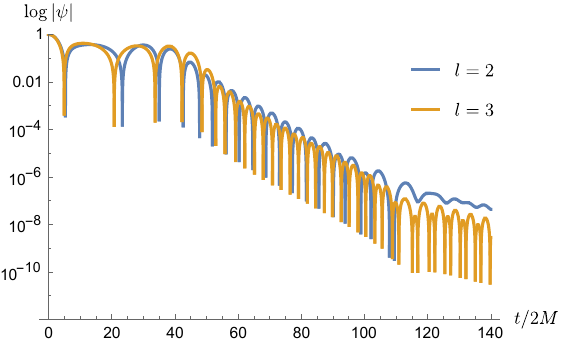}
}
\subfigure[Dark matter spike, $r_\text{b}=4$]{\label{fig:c}
\includegraphics[width=.65 \columnwidth]{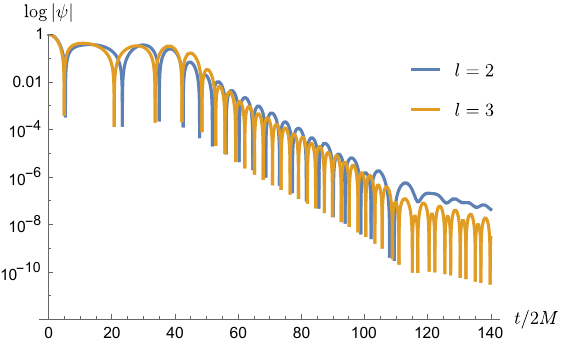}
}
\caption{The time evolutions of quasinormal modes in the axial gravitational perturbation with the different angular quantum number at the galactic center of M87. The parameters we used are $M_\text{BH}=1/2, \gamma=1, \alpha_\gamma=0.1, \rho_0=6.9 \times 10^6 M_\odot/\text{kpc}^3, r_0=91.2\text{kpc}$.}
\label{f3}
\end{figure*}
\begin{table*}[t!]
\setlength{\abovecaptionskip}{0.2cm}
\setlength{\belowcaptionskip}{0.1cm}
\setlength{\tabcolsep}{12pt}  
\renewcommand{\arraystretch}{1.2}
\centering
\caption{The fundamental quasinormal modes from the time evolutions of black holes in the axial gravitational perturbation (Fig. \ref{f3}) at the galactic center of M87 using the Prony method.}
\begin{tabular}{ccc}
\hline \hline
Types of BHs                               & $l=2$                          & $l=3$                                \\
 \hline
Schwarzschild black hole                                    & 0.7487774752 - 0.1784811211 i & 1.1916271155 - 0.1855550572 i  \\
Dark matter spike, $r_\text{b}=2$                    & 0.7487752996 - 0.1784807187 i & 1.1916235370 - 0.1855566877 i  \\
Dark matter spike, $r_\text{b}=4$                    & 0.7486766175 - 0.1784624681 i & 1.1914610882 - 0.1856305616 i  \\
\hline \hline
\end{tabular}
\label{t2}
\end{table*}

\section{Numerical results}\label{s4}
In this section, we mainly investigate the impacts of dark matter spike on the quasinormal modes of the black holes at the galactic center of M87. Firstly, to ensure the accuracy of our calculations, we utilize the Prony method and the continued fraction method to compute the quasinormal frequencies (QNF) of the Schwarzschild black hole. The results were compared with the  Leaver's results reported in Ref. \cite{Leaver:1985ax,Iyer:1986nq} and are presented in Table \ref{t1}. Overall, the numerical results obtained from the Prony method and the continued fraction method are in good agreement with the Leaver's results. Where, the calculation results obtained using the Prony method are accurate to the third decimal place. Thus, the Prony method can serve as an effective method for a rough estimation of QNF. On the other hand, the continued fraction method yields results that are nearly identical to the Leaver's results. Therefore, in this work, we mainly employ these two methods to evaluate the impacts of dark matter on the quasinormal modes of black holes. For the convenience of discussion, all these calculation results in the following subsections are presented with ten significant figures.

\subsection{Quasinormal modes of the black holes in dark matter spike at the galactic center of M87}\label{s62}
In this subsection, under the axial gravitational perturbation, we in-depth investigate the quasinormal modes of the black holes in dark matter spike at the galactic center of M87. For the quasinormal mode, its study not only aids in analyzing the stability of black holes but also enhances our understanding of the gravitational wave signals. Therefore, we first use the time-domain integration method to give the time evolution of the black hole perturbation in the dark matter spike and compare it with the Schwarzschild black hole. In Fig. \ref{f3}, we present the time evolution of the black hole perturbation under different angular quantum numbers. These time evolutions for the black hole perturbation generally occurs in three phases, with the second phase being the quasinormal mode. Here, we focus on the quasinormal modes, as they may carry crucial information about the properties of the black hole, including its mass, spin, and interaction with surrounding matter. In addition, in the stage of the quasinormal modes, the ringing value ($\log |\psi|$) of the mode $l=2$ is slightly larger than that of the mode $l=3$ at the same time. And these values of the $\log |\psi|$ under the axial gravitational perturbation are approximately between $10^{-7}$ and $10^{-2}$. In contrast, according to the recent study by Daghigh et al. \cite{Daghigh:2023ixh}, which focuses on scalar field perturbation, the corresponding QNMs range from $10^{-20}$ to $10^{-5}$. Our numerical simulations show that under the impacts of a DM spike, the QNMs for gravitational perturbations exhibit significantly higher detectability than those for scalar field perturbation. This suggests that gravitational perturbations are more sensitive to the dark matter environment, serving as a more promising observational window for testing the DM spike hypothesis. Consequently, our results may provide a theoretical basis for validating DM distribution models through future gravitational wave observations. However, the time evolution of the black hole perturbation in dark matter spike is very similar to the Schwarzschild black hole. It is difficult to distinguish the black holes in dark matter spike from the Schwarzschild black hole in terms of their time evolutions. Nevertheless, we can distinguish them using another physical quantity, that is quasinormal frequency, which can be quantified to provide a more precise comparison. Therefore, here we extract the quasinormal frequencies from their time evolutions of the black hole perturbation using the Prony method. The numerical results are shown in Table \ref{t2}. From these values in the mode $l=2$, it can be observed that the real and imaginary parts of the quasinormal frequencies for black holes in dark matter spike are slightly smaller than those for the Schwarzschild black hole. The variation of the mode $l=3$ is basically similar to the mode $l=2$. Therefore, the quasinormal frequency can serve as a distinguishing factor between the black holes in dark matter spike and Schwarzschild black hole.

\begin{table*}[t!]
\setlength{\abovecaptionskip}{0.2cm}
\setlength{\belowcaptionskip}{0.2cm}
\setlength{\tabcolsep}{6pt}  
\renewcommand{\arraystretch}{1.5}
\centering
\caption{The fundamental quasinormal modes ($n=0, l=2$) of the axial gravitational perturbation as a function of the halo density $\rho_0 (10^6M_\odot/\text{kpc}^3)$ at the fixed radius $r_0=91.2 \text{kpc}$ calculated by the continued fraction method. The corresponding quasinormal mode of Schwarzschild black hole is $\omega_\text{Sch}  =0.7473433688 - 0.1779246314 i, M_\text{BH}=1/2$.}
\scalebox{.6}{
\fontsize{9}{14}\selectfont
\begin{tabular}{ccccccc}
\hline \hline 
  & \multicolumn{3}{c}{$r_\text{b}=2$}                                                                                                            & \multicolumn{3}{c}{$r_\text{b}=4$}                                                                                                            \\
\hline
$\rho_0$         & $\gamma=0$                          & $\gamma=1/2$                               & $\gamma=1$                               & $\gamma=0$                               & $\gamma=1/2$                               & $\gamma=1$                               \\
\hline 
6.9                  & 0.7473433356 - 0.1779246235 i & 0.7473431294 - 0.1779245744 i & 0.7473411011 - 0.1779240915  i & 0.7473418124 - 0.1779242608 i & 0.7473322067 - 0.1779219739 i & 0.7472382445 - 0.1778996038 i  \\
100                 & 0.7473433039 - 0.1779246159 i & 0.7473428548 - 0.1779245090 i & 0.7473378400 - 0.1779233151 i & 0.7473403320 - 0.1779239084 i & 0.7473194098 - 0.1779189273 i & 0.7470873154 - 0.1778636712 i  \\
1000               & 0.7473432534 - 0.1779246039 i & 0.7473423763 - 0.1779243951 i & 0.7473314581 - 0.1779217957 i & 0.7473379686 - 0.1779233457 i & 0.7472971177 - 0.1779136201 i & 0.7467927672 - 0.1777935463 i  \\
\hline \hline 
\end{tabular}}
\label{t3}
\end{table*}
\begin{table*}[t!]
\setlength{\abovecaptionskip}{0.2cm}
\setlength{\belowcaptionskip}{0.2cm}
\setlength{\tabcolsep}{4.5pt}  
\renewcommand{\arraystretch}{1.5}
\centering
\caption{The fundamental quasinormal modes ($n=0, l=3$) of the axial gravitational perturbation as a function of the halo density $\rho_0 (10^6M_\odot/\text{kpc}^3)$ at the fixed radius $r_0=91.2 \text{kpc}$ calculated by the continued fraction method. The corresponding quasinormal mode of Schwarzschild black hole is $\omega_\text{Sch}  =1.198886577 - 0.1854060959 i, M_\text{BH}=1/2$.}
\scalebox{.6}{
\fontsize{10}{14}\selectfont
\begin{tabular}{ccccccc}
\hline \hline 
  & \multicolumn{3}{c}{$r_\text{b}=2$}                                                                                                            & \multicolumn{3}{c}{$r_\text{b}=4$}                                                                                                            \\
\hline
$\rho_0$         & $\gamma=0$                          & $\gamma=1/2$                               & $\gamma=1$                               & $\gamma=0$                               & $\gamma=1/2$                               & $\gamma=1$                               \\
\hline 
6.9                  & 1.198886524 - 0.1854060876 i & 1.198886193 - 0.1854060365 i & 1.198882939 - 0.1854055333 i & 1.198884080 - 0.1854057098 i & 1.198868671 - 0.1854033267 i & 1.198717937 - 0.1853800159 i  \\
100                 & 1.198886473 - 0.1854060798 i & 1.198885752 - 0.1854059684 i & 1.198877708 - 0.1854047243 i & 1.198881705 - 0.1854053425 i & 1.198848142 - 0.1854001520 i & 1.198475817 - 0.1853425724 i  \\
1000               & 1.198886392 - 0.1854060673 i & 1.198884985 - 0.1854058497 i & 1.198867470 - 0.1854031410 i & 1.198877914 - 0.1854047562 i & 1.198812381 - 0.1853946216 i & 1.198003303 - 0.1852694988 i  \\
\hline \hline 
\end{tabular}}
\label{t4}
\end{table*}

On the other hand, we also examined the impacts of the different dark matter parameters (i.e. $\rho_0$, $\gamma$, $r_\text{b}$) on the quasinormal modes of black holes and compared these results with those for Schwarzschild black holes. The numerical results are presented in Tables \ref{t3} - \ref{t4}, and the numerical method is the continued fraction method. From these results, we find that in the Schwarzschild black hole, both the real part and the absolute value of imaginary part of the quasinormal frequencies are greater than that of the black holes in dark matter spike. This indicates that the presence of dark matter spike effects the quasinormal frequencies of the black holes, leading to a decrease in both the real and imaginary parts. Specifically, the dark matter spike significantly affects the damping effect of the black holes, resulting in a longer damping time and a lower frequency for the quasinormal modes. In Table \ref{t3}, for the density of dark matter spike in the mode ($l=2$), regardless of the full relativity ($r_\text{b}=2$) or the Newtonian approximation ($r_\text{b}=4$), both the real part and the absolute value of the imaginary part of the quasinormal frequencies decrease with the increasing of the dark matter density $\rho_0$. Similarly, for the power-law index $\gamma$, the absolute values of both the real and imaginary parts of the quasinormal frequencies decrease with the increasing of the power-law index. The variations of the mode $l=3$ in Table \ref{t4} are similar to the mode $l=2$. In the near future, these results may provide some help in characterizing the impacts of dark matter spike on the quasinormal modes of the black holes. In the next subsection, for the impacts of dark matter spike on the quasinormal modes, we will explore its detectability from the perspective of space-based detectors.

\begin{table*}[t!]
\setlength{\abovecaptionskip}{0.2cm}
\setlength{\belowcaptionskip}{0.2cm}
\setlength{\tabcolsep}{10pt}  
\renewcommand{\arraystretch}{1.2}
\centering
\caption{The relative deviations of the frequency $f_{200}$ and damping time $\tau_{200}$ in dark matter spike of M87 with different power-law index $\gamma$ using the continued fraction method. The main calculation parameters are $M_\text{BH}=1/2, l=2,  \alpha_\gamma=0.1$, $\rho_0=6.9 \times 10^{6}$ $M_{\bigodot}/\text{kpc}^{3}$, $r_0=91.2$ $\text{kpc}$.}
\begin{tabular}{ccccccccc}
\hline \hline
  & \multicolumn{3}{c}{$r_\text{b}=2$}                                                                                                            & \multicolumn{3}{c}{$r_\text{b}=4$}     \\
 \hline
                                    & $\gamma=0 $            & $\gamma=1/2 $                & $\gamma=1 $               & $\gamma=0  $           & $\gamma=1/2$             & $\gamma=1$              \\ 
   \hline
$\delta f_{200}$       & 4.4424 $\times 10^{-8}$ & 3.2033 $\times 10^{-7}$      & 3.0343 $\times 10^{-6}$    & 2.0825 $\times 10^{-6}$    & 1.4935 $\times 10^{-5}$    & 1.4066 $\times 10^{-4}$       \\
$\delta \tau_{200}$  & 4.4401 $\times 10^{-8}$ & 3.2036 $\times 10^{-7}$      & 3.0344 $\times 10^{-6}$    & 2.0829 $\times 10^{-6}$    & 1.4936 $\times 10^{-5}$    & 1.4068 $\times 10^{-4}$        \\ 
\hline \hline
\end{tabular}
\label{t5}
\end{table*}

\subsection{Detectability of the impacts of dark matter spike on the quasinormal modes}\label{s62}
In the previous subsection, we investigated the quasinormal modes of the black holes in dark matter spike at the galactic center of M87, and examined the impacts of the different dark matter parameters on the quasinormal frequencies. Therefore, in this subsection, we focus on the possibility of using space-based detector to detect the impacts of dark matter spike on the quasinormal modes of the black holes. Firstly, under the axial gravitational perturbation, the ringdown waveforms of the gravitational waves in Ref. \cite{Berti:2005ys} is giving by
\begin{equation}
\begin{aligned}
h(t)&=h_+(t)+ih_\times (t)\\
&=\frac{M_z}{D_r}\sum_{lmn}A_{lmn}e^{i(f_{lmn}t+\phi_{lmn})e^{-t/\tau_{lmn}}}S_{lmn},
\label{e47}
\end{aligned}
\end{equation}
where $M_z$, $D_r$, $A_{lmn}$, $\phi_{lmn}$ respectively represent the red-shifted mass of the black holes, the luminosity distance to the source, the amplitude of the quasinormal modes, the phase coefficient, and $S_{lmn}$ denotes the spheroidal harmonics with spin weight 2. The frequency $f_{lmn}^{\text{DM}}$ and the damping time $\tau_{lmn}^{\text{DM}}$ of the quasinormal modes in dark matter spike can be expressed as
\begin{equation}
2\pi f_{lmn}^{\text{DM}}=\text{Re}(\omega_{lmn}^{\text{DM}}), \quad \tau_{lmn}^{\text{DM}}=-1/\text{Im}(\omega_{lmn}^{\text{DM}}),
\label{e48}
\end{equation}
where, $\omega_{lmn}^{\text{DM}}$ represents the quasinormal modes of the black holes under the axial gravitational perturbation at the galactic center of M87. As is known to us all, the ringdown waveforms of the gravitational waves can be derived from quadrupole radiation ($l=2$), with the $l=2$ and $l=3$ modes expected to dominate the gravitational wave signal \cite{Berti:2005ys,Berti:2009kk}. For the two modes $l=2$ and $l=3$, we select $(l,m,n)=(2,0,0)$ as the dominant mode, as it has the longest damping time, i.e., the smallest imaginary part of the quasinormal frequency (see Tables \ref{t3} and \ref{t4}). Besides, dark matter as an additional source to the vacuum black hole, it will generate the deviations from the Schwarzschild black hole. Therefore, following Ref. \cite{Zhang:2022roh}, the frequency and damping time of the black holes in dark matter spike can be described as
\begin{equation}
f_{lmn}^\text{DM}=f_{lmn}^{\text{Sch}}(1+\delta f_{lmn}),\quad \tau_{lmn}^\text{DM}=\tau_{lmn}^{\text{Sch}}(1+\delta \tau_{lmn}),
\label{e49}
\end{equation}
where, $f_{lmn}^{\text{Sch}}, \tau_{lmn}^{\text{Sch}}$ are the frequency and damping time of the quasinormal modes corresponding to Schwarzschild black hole, while $\delta f_{lmn},\delta \tau_{lmn}$ are the relative deviations to the case of the Schwarzschild black hole. 

Our numerical results obtained from the continued fraction method in Table \ref{t5} indicate that whether under the full relativity or Newtonian approximation, the relative deviations of quasinormal frequency and the damping time both increase with the increasing of the power-law index $\gamma$. The relative deviation calculated from Newtonian approximation is larger than that of the full relativity. Where, the relative deviation for the frequency is approximately $4.4424 \times10^{-8} \sim  1.4066\times10^{-4}$ and $4.4401\times10^{-8} \sim  1.4068\times10^{-4}$ for the damping time. Our these numerical results reveals that the modulation of black hole QNMs induced by the DM spike can reach the order of $10^{-4}$. This represents a significant refinement over our previous findings in Ref. \cite{Liu:2024xcd}, where the maximum impact was estimated at only $10^{-5}$ with parameter $\gamma=1$ due to the inherent limitations of the WKB and Prony methods. By employing the high-precision continued fraction method, our results show that the impacts of the DM spike are actually an order of magnitude larger than previously anticipated for the same parameter range ($\gamma = 0, 1/2, 1$). This finding suggests that lower-order approximations may underestimate the corrective effect of DM spike on the black hole QNMs. Consequently, the high-precision results provided in this work may establish a more reliable and valuable theoretical foundation for future efforts to detect dark matter modulation effects. Meanwhile, we also note that Y. Zhao et al. \cite{Zhao:2023tyo} investigated the impacts of dark matter spike on the black hole at the center of the Milky Way (Sgr A*). In comparison, our results indicate that the dark-matter-induced corrections to the quasinormal modes of M87* may be smaller by about two orders of magnitude relative to those of Sgr A*. This difference highlights that, when computing quasinormal modes with high precision, the details of the physical model and its parameter choices could have a certain impact on the final magnitude of the results. While the impacts of dark matter spike of M87 may be modest, which could still leave detectable imprints on the emitted gravitational waves. Therefore, it is necessary to compare these relative deviations with the results of space-based detectors. Currently, C. Shi et al. \cite{Shi:2019hqa} showed that the TianQin observatory can detect the maximum relative deviations in frequency and damping time, approximately 0.23\% and 1.5\%, respectively. To be specific, considering the ideal case of the LISA and TianQin detectors, the relative deviations $\delta f$ and $\delta \tau$ are approximately constrained within $0.000075< \delta f <0.0023$ and $0.00047<\delta \tau<0.015$, respectively. Upon converting to deviations in quasinormal frequencies, we obtain the following limiting conditions of the detection:
\begin{equation}
\begin{aligned}
0.7473994196<\text{Re}(\omega_{200})<0.7490622585,
\label{e50}
\end{aligned}
\end{equation}
and
\begin{equation}
\begin{aligned}
0.1752952033<-\text{Im}(\omega_{200})<0.1778410461.
\label{e51}
\end{aligned}
\end{equation}
In other words, the Tianqin Observatory is currently only able to detect relative deviations of the order of $10^{-3}$ at most. This result closely approximates our predictions although there are still some gaps. In the near future, with ongoing advancements in observational technology, we believe the next generation of detectors will be capable of detecting even smaller deviations. We hope these research results will assist in the indirect detection of dark matter from the perspective of the gravitational waves.

\section{Summary}\label{s5}
In this work, we first obtain the analytical solutions of black holes in dark matter spike by solving the TOV equation. Then, combined with the mass model of M87, we investigate the gravitational wave ringdown waveforms of the black holes in dark matter spike under the axial gravitational perturbation, that is quasinormal modes, and compare them with the Schwarzschild black hole. Meanwhile, we also examine the impacts of the different dark matter parameters on the quasinormal modes of the black holes. Finally, the detectability for the space-based detectors to detect these impacts of dark matter spike on the quasinormal modes is explored. Our results are as follows:

(1) With the help of Eq. (\ref{e12}), after fixing the power-law index $\gamma$, the analytical solutions of the black holes in dark matter spike can be obtained through the TOV equation. Although the analytical solution for the black hole metric can only be obtained under specific parameter conditions, this does not hinder our ability to study our problems. These black hole analytical solutions can be found in Appendix \ref{A}.

(2) Combined with the mass model of M87, the time evolutions of the quasinormal modes for black holes in dark matter spike are presented in Fig. \ref{f3}.  And then the quasinormal frequencies are investigated using the Prony method and continued fraction method in Tables \ref{t1} and \ref{t2}. Both these two methods show that the real and imaginary parts of the quasinormal frequencies for black holes in dark matter spike are slightly smaller than those for the Schwarzschild black hole. This indicates that the presence of dark matter spike effects the quasinormal frequencies of the black holes, leading to a decrease in both the real and imaginary parts. Specifically, the dark matter spike significantly affects the damping effect of the black holes, resulting in a longer damping time and a lower frequency for the quasinormal modes.

(3) The impacts of the different dark matter parameters on the quasinormal modes of the black holes are examined using the continued fraction method in Tables \ref{t3} and \ref{t4}. In the Schwarzschild black hole, both the real part and the absolute value of imaginary part of the quasinormal frequencies are greater than that of the black holes in dark matter spike. This indicates that the presence of dark matter spike effects the quasinormal frequencies of the black holes, leading to a decrease in both the real and imaginary parts. Regardless of the full relativity or the Newtonian approximation, both the real part and the absolute value of the imaginary part of the quasinormal frequencies decrease with the increasing of the dark matter density $\rho_0$. Besides, the absolute values of both the real and imaginary parts of the quasinormal frequencies decrease with the increasing of the power-law index $\gamma$. In the near future, these results may provide some help in characterizing the impacts of dark matter spike on the quasinormal modes of the black holes.

(4) The detectability for the space-based detectors to detect these impacts of dark matter spike on the quasinormal modes is explored in Table \ref{t5}. Whether under the full relativity or the Newtonian approximation, the relative deviations of quasinormal frequency and the damping time both increase with the increasing of the power-law index $\gamma$. The relative deviations calculated from Newtonian approximation is larger than that of the full relativity. Where, the relative deviations for these frequencies are approximately $4.4424 \times10^{-8} \sim  1.4066\times10^{-4}$ and $4.4401\times10^{-8} \sim  1.4068\times10^{-4}$ for the damping time. From these numerical results,  the impacts of dark matter spike on the quasinormal modes of the black holes can reach up to the order of $10^{-4}$ at most. However, the Tianqin Observatory is currently only able to detect relative deviations of the order of $10^{-3}$ at most. This result closely approximates our predictions although there are still some gaps. In the near future, with ongoing advancements in observational technology, we believe the next generation of detectors will be capable of detecting even smaller deviations. We hope these results will assist in the indirect detection of dark matter from the perspective of the gravitational waves.

\begin{acknowledgments}
Dong Liu would like to acknowledge the anonymous referee for their constructive report, which has significantly improved this paper. 
This work was partly supported by the Guizhou Provincial Basic Research Program (Natural Science) under Grant No.QN[2025]310 and the Project of Guizhou Provincial Department of Science and Technology under Grant No.CXTD[2025]030. This work also was funded by the National Natural Science Foundation of China (No. 12505064), Guizhou Provincial Basic Research Program (Natural Science) Youth Guidance Program (No. QN [2025] 365) and the project of Young Scientific and Technical Talents Devel opment of Education Department of Guizhou Province under Grant [2024] 79.
\end{acknowledgments}

\onecolumngrid
\appendix
\section{Black hole metrics $g(r)$ and $f(r)$ for the different power-law indexes} \label{A}
In this work, the background metric of black holes in dark matter spike is characterized by the line element
\begin{equation}
\begin{aligned}
ds^{2}=-f(r)dt^{2}+ g(r)^{-1}dr^{2}+r^{2}(d\theta ^{2}+\sin^{2}\theta d\phi ^{2}),
\end{aligned}
\end{equation}
where,  we can always choose metric function $g(r)=1-2m(r)/r$, and it reads
{\small
\begin{equation}
g(r)=1-\frac{2M_\text{BH}}{r}-\frac{8 \pi \rho_\text{sp}}{r} \left( 
\frac{6 r_\text{b}^3 \left(R_\text{sp}/r_\text{b}\right)^{\gamma_\text{sp}}}{\gamma_\text{sp}(\gamma_\text{sp}-1)(\gamma_\text{sp}-2)(\gamma_\text{sp}-3)} +
\left( \frac{r^3}{3-\gamma_\text{sp}} + \frac{3 r^2 r_\text{b}}{\gamma_\text{sp}-2} - \frac{3 r r_\text{b}^2}{\gamma_\text{sp}-1} + \frac{r_\text{b}^3}{\gamma_\text{sp}} \right) 
\left( \frac{R_\text{sp}}{r} \right)^{\gamma_\text{sp}} 
\right).
\end{equation}
}

For the metric function $f(r)$, it can be obtained from TOV equation with the help of $g(r)$, that is
\begin{equation}
\begin{aligned}
\frac{f'(r)}{f(r)}=-\frac{1}{r}+\frac{1}{r-2(M_\text{BH}+a_1r^{3-\gamma _{\text{sp}}}+a_2r^{2-\gamma _{\text{sp}}}+a_3r^{1-\gamma _{\text{sp}}}+a_4r^{-\gamma _{\text{sp}}}+a_5)},
\label{ea3}
\end{aligned}
\end{equation}
where, the coefficients $a_1$, $a_2$, $a_3$, $a_4$, $a_5$ are as follows:
\begin{equation}
\begin{aligned}
&a_1=-\frac{4 \pi \rho_{\text{sp}} R_{\text{sp}}^{\gamma_{\text{sp}}}}{\gamma_{\text{sp}}-3}, \quad a_2=\frac{12 \pi \rho_{\text{sp}} r_\text{b} R_{\text{sp}}^{\gamma_{\text{sp}}}}{\gamma_{\text{sp}}-2}, \quad a_3=-\frac{12 \pi \rho_{\text{sp}} r_\text{b}^2 R_{\text{sp}}^{\gamma_{\text{sp}}}}{\gamma_{\text{sp}}-1}, \\
&a_4=\frac{4 \pi \rho_{\text{sp}} r_\text{b}^3 R_{\text{sp}}^{\gamma_{\text{sp}}}}{\gamma_{\text{sp}}}, \quad a_5=\frac{24 \pi \rho_{\text{sp}} r_\text{b}^3 (R_{\text{sp}}/r_\text{b})^{\gamma_{\text{sp}}}}{\gamma_{\text{sp}}(\gamma_{\text{sp}}-1)(\gamma_{\text{sp}}-2)(\gamma_{\text{sp}}-3)}.
\end{aligned}
\end{equation}

\textbf{Case I: Power-law index $\gamma=0$, and then $\gamma_\text{sp}=9/4$.}

After integrating Eq. (\ref{ea3}) using $\gamma_\text{sp}=9/4$, the final analytic expression for the black hole metric $f_1(r)$ is
\begin{equation}
\begin{aligned}
f_1(r)=(\frac{r_\text{b}-2M_\text{BH}}{r})\times \exp \left[4\times [\mathcal{F}_1(r)-\mathcal{F}_1(r_\text{b})] \right],\\
\end{aligned}
\end{equation}
where, $\mathcal{F}_1(r)$ and $\mathcal{F}_1(r_\text{b})$ are
\begin{equation}
\begin{aligned}
\mathcal{F}_1(r)= \sum_{x_i} \frac{ \log[(r^{1/4}-x_i)]\times  x_i^9}{-8a_3-16a_2x_i^4-18a_5x_i^5-18M_\text{BH}x_i^5-24a_1x_i^8+13x_i^9},\\
\mathcal{F}_1(r_\text{b})= \sum_{x_i} \frac{ \log[(r_\text{b}^{1/4}-x_i)]\times  x_i^9}{-8a_3-16a_2x_i^4-18a_5x_i^5-18M_\text{BH}x_i^5-24a_1x_i^8+13x_i^9},\\
\end{aligned}
\end{equation}
while $x_i$ is the root of the following equation
\begin{equation}
\begin{aligned}
2a_4+2a_3x^4+2a_2x^8+2a_5x^9+2M_\text{BH}x^9+2a_1x^{12}-x^{13}=0.
\end{aligned}
\end{equation}

\textbf{Case II: Power-law index $\gamma=1/2$, and then $\gamma_\text{sp}=16/7$.}

After integrating Eq. (\ref{ea3}) using $\gamma_\text{sp}=16/7$, the final analytic expression for the black hole metric $f_2(r)$ is
\begin{equation}
\begin{aligned}
f_2(r)=(\frac{r_\text{b}-2M_\text{BH}}{r})\times \exp \left[7\times [\mathcal{F}_2(r)-\mathcal{F}_2(r_\text{b})] \right],
\end{aligned}
\end{equation}
where, $\mathcal{F}_2(r)$ and $\mathcal{F}_2(r_\text{b})$ are
\begin{equation}
\begin{aligned}
\mathcal{F}_2(r)= \sum_{x_i} \frac{ \log[(r^{1/7}-x_i)]\times  x_i^{16}}{-14a_3-28a_2x_i^7-32a_5x_i^9-32M_\text{BH}x_i^9-42a_1x_i^{14}+23x_i^{16}},\\
\mathcal{F}_2(r_\text{b})= \sum_{x_i} \frac{ \log[(r_\text{b}^{1/7}-x_i)]\times  x_i^{16}}{-14a_3-28a_2x_i^7-32a_5x_i^9-32M_\text{BH}x_i^9-42a_1x_i^{14}+23x_i^{16}},
\end{aligned}
\end{equation}
while $x_i$ is the root of the following equation
\begin{equation}
\begin{aligned}
2a_4+2a_3x^7+2a_2x^{14}+2a_5x^{16}+2M_\text{BH}x^{16}+2a_1x^{21}-x^{23}=0.
\end{aligned}
\end{equation}

\textbf{Case III: Power-law index $\gamma=1$, and then $\gamma_\text{sp}=7/3$.}

After integrating Eq. (\ref{ea3}) using $\gamma_\text{sp}=7/3$, the final analytic expression for the black hole metric $f_3(r)$ is
\begin{equation}
f_3(r)=(\frac{r_\text{b}-2M_\text{BH}}{r})\times \exp \left[\frac{3}{2}\times [\mathcal{F}_3(r)-\mathcal{F}_3(r_\text{b})] \right],
\end{equation}
where, $\mathcal{F}_3(r)$ and $\mathcal{F}_3(r_\text{b})$ are
\begin{equation}
\begin{aligned}
&\mathcal{F}_3(r)= \sum_{x_i} \frac{ \log[(r^{1/3}-x_i)]\times  x_i^7}{-3a_3-6a_2x_i^3-7a_5x_i^4-7M_\text{BH}x_i^4-9a_1x_i^6+5x_i^7},\\
&\mathcal{F}_3(r_\text{b})= \sum_{x_i} \frac{ \log[(r_\text{b}^{1/3}-x_i)]\times  x_i^7}{-3a_3-6a_2x_i^3-7a_5x_i^4-7M_\text{BH}x_i^4-9a_1x_i^6+5x_i^7},
\end{aligned}
\end{equation}
while $x_i$ is the root of the following equation
\begin{equation}
\begin{aligned}
2a_4+2a_3x^3+2a_2x^6+2a_5x^7+2M_\text{BH}x^7+2a_1x^9-x^{10}=0.
\end{aligned}
\end{equation}

\twocolumngrid
\hypersetup{urlcolor=blue}
\bibliographystyle{apsrev4-1}
\bibliography{ref}

\begin{thebibliography}{67}%
\makeatletter
\providecommand \@ifxundefined [1]{%
 \@ifx{#1\undefined}
}%
\providecommand \@ifnum [1]{%
 \ifnum #1\expandafter \@firstoftwo
 \else \expandafter \@secondoftwo
 \fi
}%
\providecommand \@ifx [1]{%
 \ifx #1\expandafter \@firstoftwo
 \else \expandafter \@secondoftwo
 \fi
}%
\providecommand \natexlab [1]{#1}%
\providecommand \enquote  [1]{``#1''}%
\providecommand \bibnamefont  [1]{#1}%
\providecommand \bibfnamefont [1]{#1}%
\providecommand \citenamefont [1]{#1}%
\providecommand \href@noop [0]{\@secondoftwo}%
\providecommand \href [0]{\begingroup \@sanitize@url \@href}%
\providecommand \@href[1]{\@@startlink{#1}\@@href}%
\providecommand \@@href[1]{\endgroup#1\@@endlink}%
\providecommand \@sanitize@url [0]{\catcode `\\12\catcode `\$12\catcode
  `\&12\catcode `\#12\catcode `\^12\catcode `\_12\catcode `\%12\relax}%
\providecommand \@@startlink[1]{}%
\providecommand \@@endlink[0]{}%
\providecommand \url  [0]{\begingroup\@sanitize@url \@url }%
\providecommand \@url [1]{\endgroup\@href {#1}{\urlprefix }}%
\providecommand \urlprefix  [0]{URL }%
\providecommand \Eprint [0]{\href }%
\providecommand \doibase [0]{http://dx.doi.org/}%
\providecommand \selectlanguage [0]{\@gobble}%
\providecommand \bibinfo  [0]{\@secondoftwo}%
\providecommand \bibfield  [0]{\@secondoftwo}%
\providecommand \translation [1]{[#1]}%
\providecommand \BibitemOpen [0]{}%
\providecommand \bibitemStop [0]{}%
\providecommand \bibitemNoStop [0]{.\EOS\space}%
\providecommand \EOS [0]{\spacefactor3000\relax}%
\providecommand \BibitemShut  [1]{\csname bibitem#1\endcsname}%
\let\auto@bib@innerbib\@empty
\bibitem [{\citenamefont {Aghanim}\ \emph {et~al.}(2020)\citenamefont {Aghanim}
  \emph {et~al.}}]{Planck:2018vyg}%
  \BibitemOpen
  \bibfield  {author} {\bibinfo {author} {\bibfnamefont {N.}~\bibnamefont
  {Aghanim}} \emph {et~al.} (\bibinfo {collaboration} {Planck}),\ }\href
  {\doibase 10.1051/0004-6361/201833910} {\bibfield  {journal} {\bibinfo
  {journal} {Astron. Astrophys.}\ }\textbf {\bibinfo {volume} {641}},\ \bibinfo
  {pages} {A6} (\bibinfo {year} {2020})},\ \bibinfo {note} {[Erratum:
  Astron.Astrophys. 652, C4 (2021)]},\ \Eprint
  {http://arxiv.org/abs/1807.06209} {arXiv:1807.06209 [astro-ph.CO]}
  \BibitemShut {NoStop}%
\bibitem [{\citenamefont {Freese}(2009)}]{Freese:2008cz}%
  \BibitemOpen
  \bibfield  {author} {\bibinfo {author} {\bibfnamefont {K.}~\bibnamefont
  {Freese}},\ }\href {\doibase 10.1051/eas/0936016} {\bibfield  {journal}
  {\bibinfo  {journal} {EAS Publ. Ser.}\ }\textbf {\bibinfo {volume} {36}},\
  \bibinfo {pages} {113} (\bibinfo {year} {2009})},\ \Eprint
  {http://arxiv.org/abs/0812.4005} {arXiv:0812.4005 [astro-ph]} \BibitemShut
  {NoStop}%
\bibitem [{\citenamefont {Navarro}\ \emph {et~al.}(1996)\citenamefont
  {Navarro}, \citenamefont {Frenk},\ and\ \citenamefont
  {White}}]{Navarro:1995iw}%
  \BibitemOpen
  \bibfield  {author} {\bibinfo {author} {\bibfnamefont {J.~F.}\ \bibnamefont
  {Navarro}}, \bibinfo {author} {\bibfnamefont {C.~S.}\ \bibnamefont {Frenk}},
  \ and\ \bibinfo {author} {\bibfnamefont {S.~D.~M.}\ \bibnamefont {White}},\
  }\href {\doibase 10.1086/177173} {\bibfield  {journal} {\bibinfo  {journal}
  {Astrophys. J.}\ }\textbf {\bibinfo {volume} {462}},\ \bibinfo {pages} {563}
  (\bibinfo {year} {1996})},\ \Eprint {http://arxiv.org/abs/astro-ph/9508025}
  {arXiv:astro-ph/9508025} \BibitemShut {NoStop}%
\bibitem [{\citenamefont {Clowe}\ \emph {et~al.}(2006)\citenamefont {Clowe},
  \citenamefont {Bradac}, \citenamefont {Gonzalez}, \citenamefont {Markevitch},
  \citenamefont {Randall}, \citenamefont {Jones},\ and\ \citenamefont
  {Zaritsky}}]{Clowe:2006eq}%
  \BibitemOpen
  \bibfield  {author} {\bibinfo {author} {\bibfnamefont {D.}~\bibnamefont
  {Clowe}}, \bibinfo {author} {\bibfnamefont {M.}~\bibnamefont {Bradac}},
  \bibinfo {author} {\bibfnamefont {A.~H.}\ \bibnamefont {Gonzalez}}, \bibinfo
  {author} {\bibfnamefont {M.}~\bibnamefont {Markevitch}}, \bibinfo {author}
  {\bibfnamefont {S.~W.}\ \bibnamefont {Randall}}, \bibinfo {author}
  {\bibfnamefont {C.}~\bibnamefont {Jones}}, \ and\ \bibinfo {author}
  {\bibfnamefont {D.}~\bibnamefont {Zaritsky}},\ }\href {\doibase
  10.1086/508162} {\bibfield  {journal} {\bibinfo  {journal} {Astrophys. J.
  Lett.}\ }\textbf {\bibinfo {volume} {648}},\ \bibinfo {pages} {L109}
  (\bibinfo {year} {2006})},\ \Eprint {http://arxiv.org/abs/astro-ph/0608407}
  {arXiv:astro-ph/0608407} \BibitemShut {NoStop}%
\bibitem [{\citenamefont {Barack}\ \emph {et~al.}(2019)\citenamefont {Barack}
  \emph {et~al.}}]{Barack:2018yly}%
  \BibitemOpen
  \bibfield  {author} {\bibinfo {author} {\bibfnamefont {L.}~\bibnamefont
  {Barack}} \emph {et~al.},\ }\href {\doibase 10.1088/1361-6382/ab0587}
  {\bibfield  {journal} {\bibinfo  {journal} {Class. Quant. Grav.}\ }\textbf
  {\bibinfo {volume} {36}},\ \bibinfo {pages} {143001} (\bibinfo {year}
  {2019})},\ \Eprint {http://arxiv.org/abs/1806.05195} {arXiv:1806.05195
  [gr-qc]} \BibitemShut {NoStop}%
\bibitem [{\citenamefont {Cardoso}\ and\ \citenamefont
  {Pani}(2019)}]{Cardoso:2019rvt}%
  \BibitemOpen
  \bibfield  {author} {\bibinfo {author} {\bibfnamefont {V.}~\bibnamefont
  {Cardoso}}\ and\ \bibinfo {author} {\bibfnamefont {P.}~\bibnamefont {Pani}},\
  }\href {\doibase 10.1007/s41114-019-0020-4} {\bibfield  {journal} {\bibinfo
  {journal} {Living Rev. Rel.}\ }\textbf {\bibinfo {volume} {22}},\ \bibinfo
  {pages} {4} (\bibinfo {year} {2019})},\ \Eprint
  {http://arxiv.org/abs/1904.05363} {arXiv:1904.05363 [gr-qc]} \BibitemShut
  {NoStop}%
\bibitem [{\citenamefont {Bar}\ \emph {et~al.}(2019)\citenamefont {Bar},
  \citenamefont {Blum}, \citenamefont {Lacroix},\ and\ \citenamefont
  {Panci}}]{Bar:2019pnz}%
  \BibitemOpen
  \bibfield  {author} {\bibinfo {author} {\bibfnamefont {N.}~\bibnamefont
  {Bar}}, \bibinfo {author} {\bibfnamefont {K.}~\bibnamefont {Blum}}, \bibinfo
  {author} {\bibfnamefont {T.}~\bibnamefont {Lacroix}}, \ and\ \bibinfo
  {author} {\bibfnamefont {P.}~\bibnamefont {Panci}},\ }\href {\doibase
  10.1088/1475-7516/2019/07/045} {\bibfield  {journal} {\bibinfo  {journal}
  {JCAP}\ }\textbf {\bibinfo {volume} {07}},\ \bibinfo {pages} {045} (\bibinfo
  {year} {2019})},\ \Eprint {http://arxiv.org/abs/1905.11745} {arXiv:1905.11745
  [astro-ph.CO]} \BibitemShut {NoStop}%
\bibitem [{\citenamefont {Abbott}\ \emph {et~al.}(2016)\citenamefont {Abbott}
  \emph {et~al.}}]{LIGOScientific:2016aoc}%
  \BibitemOpen
  \bibfield  {author} {\bibinfo {author} {\bibfnamefont {B.~P.}\ \bibnamefont
  {Abbott}} \emph {et~al.} (\bibinfo {collaboration} {LIGO Scientific,
  Virgo}),\ }\href {\doibase 10.1103/PhysRevLett.116.061102} {\bibfield
  {journal} {\bibinfo  {journal} {Phys. Rev. Lett.}\ }\textbf {\bibinfo
  {volume} {116}},\ \bibinfo {pages} {061102} (\bibinfo {year} {2016})},\
  \Eprint {http://arxiv.org/abs/1602.03837} {arXiv:1602.03837 [gr-qc]}
  \BibitemShut {NoStop}%
\bibitem [{\citenamefont {Akiyama}\ \emph {et~al.}(2019)\citenamefont {Akiyama}
  \emph {et~al.}}]{EventHorizonTelescope:2019dse}%
  \BibitemOpen
  \bibfield  {author} {\bibinfo {author} {\bibfnamefont {K.}~\bibnamefont
  {Akiyama}} \emph {et~al.} (\bibinfo {collaboration} {Event Horizon
  Telescope}),\ }\href {\doibase 10.3847/2041-8213/ab0ec7} {\bibfield
  {journal} {\bibinfo  {journal} {Astrophys. J. Lett.}\ }\textbf {\bibinfo
  {volume} {875}},\ \bibinfo {pages} {L1} (\bibinfo {year} {2019})},\ \Eprint
  {http://arxiv.org/abs/1906.11238} {arXiv:1906.11238 [astro-ph.GA]}
  \BibitemShut {NoStop}%
\bibitem [{\citenamefont {Akiyama}\ \emph {et~al.}(2022)\citenamefont {Akiyama}
  \emph {et~al.}}]{EventHorizonTelescope:2022apq}%
  \BibitemOpen
  \bibfield  {author} {\bibinfo {author} {\bibfnamefont {K.}~\bibnamefont
  {Akiyama}} \emph {et~al.} (\bibinfo {collaboration} {Event Horizon
  Telescope}),\ }\href {\doibase 10.3847/2041-8213/ac6675} {\bibfield
  {journal} {\bibinfo  {journal} {Astrophys. J. Lett.}\ }\textbf {\bibinfo
  {volume} {930}},\ \bibinfo {pages} {L13} (\bibinfo {year}
  {2022})}\BibitemShut {NoStop}%
\bibitem [{\citenamefont {Konoplya}\ and\ \citenamefont
  {Zhidenko}(2011{\natexlab{a}})}]{Konoplya:2011qq}%
  \BibitemOpen
  \bibfield  {author} {\bibinfo {author} {\bibfnamefont {R.~A.}\ \bibnamefont
  {Konoplya}}\ and\ \bibinfo {author} {\bibfnamefont {A.}~\bibnamefont
  {Zhidenko}},\ }\href {\doibase 10.1103/RevModPhys.83.793} {\bibfield
  {journal} {\bibinfo  {journal} {Rev. Mod. Phys.}\ }\textbf {\bibinfo {volume}
  {83}},\ \bibinfo {pages} {793} (\bibinfo {year} {2011}{\natexlab{a}})},\
  \Eprint {http://arxiv.org/abs/1102.4014} {arXiv:1102.4014 [gr-qc]}
  \BibitemShut {NoStop}%
\bibitem [{\citenamefont {Yunes}\ \emph {et~al.}(2011)\citenamefont {Yunes},
  \citenamefont {Kocsis}, \citenamefont {Loeb},\ and\ \citenamefont
  {Haiman}}]{Yunes:2011ws}%
  \BibitemOpen
  \bibfield  {author} {\bibinfo {author} {\bibfnamefont {N.}~\bibnamefont
  {Yunes}}, \bibinfo {author} {\bibfnamefont {B.}~\bibnamefont {Kocsis}},
  \bibinfo {author} {\bibfnamefont {A.}~\bibnamefont {Loeb}}, \ and\ \bibinfo
  {author} {\bibfnamefont {Z.}~\bibnamefont {Haiman}},\ }\href {\doibase
  10.1103/PhysRevLett.107.171103} {\bibfield  {journal} {\bibinfo  {journal}
  {Phys. Rev. Lett.}\ }\textbf {\bibinfo {volume} {107}},\ \bibinfo {pages}
  {171103} (\bibinfo {year} {2011})},\ \Eprint {http://arxiv.org/abs/1103.4609}
  {arXiv:1103.4609 [astro-ph.CO]} \BibitemShut {NoStop}%
\bibitem [{\citenamefont {Cardoso}\ \emph {et~al.}(2021)\citenamefont
  {Cardoso}, \citenamefont {Macedo},\ and\ \citenamefont
  {Vicente}}]{Cardoso:2020iji}%
  \BibitemOpen
  \bibfield  {author} {\bibinfo {author} {\bibfnamefont {V.}~\bibnamefont
  {Cardoso}}, \bibinfo {author} {\bibfnamefont {C.~F.~B.}\ \bibnamefont
  {Macedo}}, \ and\ \bibinfo {author} {\bibfnamefont {R.}~\bibnamefont
  {Vicente}},\ }\href {\doibase 10.1103/PhysRevD.103.023015} {\bibfield
  {journal} {\bibinfo  {journal} {Phys. Rev. D}\ }\textbf {\bibinfo {volume}
  {103}},\ \bibinfo {pages} {023015} (\bibinfo {year} {2021})},\ \Eprint
  {http://arxiv.org/abs/2010.15151} {arXiv:2010.15151 [gr-qc]} \BibitemShut
  {NoStop}%
\bibitem [{\citenamefont {Zwick}\ \emph {et~al.}(2023)\citenamefont {Zwick},
  \citenamefont {Capelo},\ and\ \citenamefont {Mayer}}]{Zwick:2022dih}%
  \BibitemOpen
  \bibfield  {author} {\bibinfo {author} {\bibfnamefont {L.}~\bibnamefont
  {Zwick}}, \bibinfo {author} {\bibfnamefont {P.~R.}\ \bibnamefont {Capelo}}, \
  and\ \bibinfo {author} {\bibfnamefont {L.}~\bibnamefont {Mayer}},\ }\href
  {\doibase 10.1093/mnras/stad707} {\bibfield  {journal} {\bibinfo  {journal}
  {Mon. Not. Roy. Astron. Soc.}\ }\textbf {\bibinfo {volume} {521}},\ \bibinfo
  {pages} {4645} (\bibinfo {year} {2023})},\ \Eprint
  {http://arxiv.org/abs/2209.04060} {arXiv:2209.04060 [gr-qc]} \BibitemShut
  {NoStop}%
\bibitem [{\citenamefont {Eda}\ \emph {et~al.}(2013)\citenamefont {Eda},
  \citenamefont {Itoh}, \citenamefont {Kuroyanagi},\ and\ \citenamefont
  {Silk}}]{Eda:2013gg}%
  \BibitemOpen
  \bibfield  {author} {\bibinfo {author} {\bibfnamefont {K.}~\bibnamefont
  {Eda}}, \bibinfo {author} {\bibfnamefont {Y.}~\bibnamefont {Itoh}}, \bibinfo
  {author} {\bibfnamefont {S.}~\bibnamefont {Kuroyanagi}}, \ and\ \bibinfo
  {author} {\bibfnamefont {J.}~\bibnamefont {Silk}},\ }\href {\doibase
  10.1103/PhysRevLett.110.221101} {\bibfield  {journal} {\bibinfo  {journal}
  {Phys. Rev. Lett.}\ }\textbf {\bibinfo {volume} {110}},\ \bibinfo {pages}
  {221101} (\bibinfo {year} {2013})},\ \Eprint {http://arxiv.org/abs/1301.5971}
  {arXiv:1301.5971 [gr-qc]} \BibitemShut {NoStop}%
\bibitem [{\citenamefont {Macedo}\ \emph {et~al.}(2013)\citenamefont {Macedo},
  \citenamefont {Pani}, \citenamefont {Cardoso},\ and\ \citenamefont
  {Crispino}}]{Macedo:2013qea}%
  \BibitemOpen
  \bibfield  {author} {\bibinfo {author} {\bibfnamefont {C.~F.~B.}\
  \bibnamefont {Macedo}}, \bibinfo {author} {\bibfnamefont {P.}~\bibnamefont
  {Pani}}, \bibinfo {author} {\bibfnamefont {V.}~\bibnamefont {Cardoso}}, \
  and\ \bibinfo {author} {\bibfnamefont {L.~C.~B.}\ \bibnamefont {Crispino}},\
  }\href {\doibase 10.1088/0004-637X/774/1/48} {\bibfield  {journal} {\bibinfo
  {journal} {Astrophys. J.}\ }\textbf {\bibinfo {volume} {774}},\ \bibinfo
  {pages} {48} (\bibinfo {year} {2013})},\ \Eprint
  {http://arxiv.org/abs/1302.2646} {arXiv:1302.2646 [gr-qc]} \BibitemShut
  {NoStop}%
\bibitem [{\citenamefont {Traykova}\ \emph {et~al.}(2021)\citenamefont
  {Traykova}, \citenamefont {Clough}, \citenamefont {Helfer}, \citenamefont
  {Berti}, \citenamefont {Ferreira},\ and\ \citenamefont
  {Hui}}]{Traykova:2021dua}%
  \BibitemOpen
  \bibfield  {author} {\bibinfo {author} {\bibfnamefont {D.}~\bibnamefont
  {Traykova}}, \bibinfo {author} {\bibfnamefont {K.}~\bibnamefont {Clough}},
  \bibinfo {author} {\bibfnamefont {T.}~\bibnamefont {Helfer}}, \bibinfo
  {author} {\bibfnamefont {E.}~\bibnamefont {Berti}}, \bibinfo {author}
  {\bibfnamefont {P.~G.}\ \bibnamefont {Ferreira}}, \ and\ \bibinfo {author}
  {\bibfnamefont {L.}~\bibnamefont {Hui}},\ }\href {\doibase
  10.1103/PhysRevD.104.103014} {\bibfield  {journal} {\bibinfo  {journal}
  {Phys. Rev. D}\ }\textbf {\bibinfo {volume} {104}},\ \bibinfo {pages}
  {103014} (\bibinfo {year} {2021})},\ \Eprint
  {http://arxiv.org/abs/2106.08280} {arXiv:2106.08280 [gr-qc]} \BibitemShut
  {NoStop}%
\bibitem [{\citenamefont {Li}\ \emph {et~al.}(2022)\citenamefont {Li},
  \citenamefont {Tang},\ and\ \citenamefont {Wu}}]{Li:2021pxf}%
  \BibitemOpen
  \bibfield  {author} {\bibinfo {author} {\bibfnamefont {G.-L.}\ \bibnamefont
  {Li}}, \bibinfo {author} {\bibfnamefont {Y.}~\bibnamefont {Tang}}, \ and\
  \bibinfo {author} {\bibfnamefont {Y.-L.}\ \bibnamefont {Wu}},\ }\href
  {\doibase 10.1007/s11433-022-1930-9} {\bibfield  {journal} {\bibinfo
  {journal} {Sci. China Phys. Mech. Astron.}\ }\textbf {\bibinfo {volume}
  {65}},\ \bibinfo {pages} {100412} (\bibinfo {year} {2022})},\ \Eprint
  {http://arxiv.org/abs/2112.14041} {arXiv:2112.14041 [astro-ph.CO]}
  \BibitemShut {NoStop}%
\bibitem [{\citenamefont {Dai}\ \emph {et~al.}(2024)\citenamefont {Dai},
  \citenamefont {Gong}, \citenamefont {Zhao},\ and\ \citenamefont
  {Jiang}}]{Dai:2023cft}%
  \BibitemOpen
  \bibfield  {author} {\bibinfo {author} {\bibfnamefont {N.}~\bibnamefont
  {Dai}}, \bibinfo {author} {\bibfnamefont {Y.}~\bibnamefont {Gong}}, \bibinfo
  {author} {\bibfnamefont {Y.}~\bibnamefont {Zhao}}, \ and\ \bibinfo {author}
  {\bibfnamefont {T.}~\bibnamefont {Jiang}},\ }\href {\doibase
  10.1103/PhysRevD.110.084080} {\bibfield  {journal} {\bibinfo  {journal}
  {Phys. Rev. D}\ }\textbf {\bibinfo {volume} {110}},\ \bibinfo {pages}
  {084080} (\bibinfo {year} {2024})},\ \Eprint
  {http://arxiv.org/abs/2301.05088} {arXiv:2301.05088 [gr-qc]} \BibitemShut
  {NoStop}%
\bibitem [{\citenamefont {Xu}\ \emph {et~al.}(2018)\citenamefont {Xu},
  \citenamefont {Hou}, \citenamefont {Gong},\ and\ \citenamefont
  {Wang}}]{Xu:2018wow}%
  \BibitemOpen
  \bibfield  {author} {\bibinfo {author} {\bibfnamefont {Z.}~\bibnamefont
  {Xu}}, \bibinfo {author} {\bibfnamefont {X.}~\bibnamefont {Hou}}, \bibinfo
  {author} {\bibfnamefont {X.}~\bibnamefont {Gong}}, \ and\ \bibinfo {author}
  {\bibfnamefont {J.}~\bibnamefont {Wang}},\ }\href {\doibase
  10.1088/1475-7516/2018/09/038} {\bibfield  {journal} {\bibinfo  {journal}
  {JCAP}\ }\textbf {\bibinfo {volume} {09}},\ \bibinfo {pages} {038} (\bibinfo
  {year} {2018})},\ \Eprint {http://arxiv.org/abs/1803.00767} {arXiv:1803.00767
  [gr-qc]} \BibitemShut {NoStop}%
\bibitem [{\citenamefont {Xu}\ \emph {et~al.}(2021)\citenamefont {Xu},
  \citenamefont {Wang},\ and\ \citenamefont {Tang}}]{Xu:2021dkv}%
  \BibitemOpen
  \bibfield  {author} {\bibinfo {author} {\bibfnamefont {Z.}~\bibnamefont
  {Xu}}, \bibinfo {author} {\bibfnamefont {J.}~\bibnamefont {Wang}}, \ and\
  \bibinfo {author} {\bibfnamefont {M.}~\bibnamefont {Tang}},\ }\href {\doibase
  10.1088/1475-7516/2021/09/007} {\bibfield  {journal} {\bibinfo  {journal}
  {JCAP}\ }\textbf {\bibinfo {volume} {09}},\ \bibinfo {pages} {007} (\bibinfo
  {year} {2021})},\ \Eprint {http://arxiv.org/abs/2104.13158} {arXiv:2104.13158
  [gr-qc]} \BibitemShut {NoStop}%
\bibitem [{\citenamefont {Cardoso}\ \emph {et~al.}(2022)\citenamefont
  {Cardoso}, \citenamefont {Destounis}, \citenamefont {Duque}, \citenamefont
  {Macedo},\ and\ \citenamefont {Maselli}}]{Cardoso:2021wlq}%
  \BibitemOpen
  \bibfield  {author} {\bibinfo {author} {\bibfnamefont {V.}~\bibnamefont
  {Cardoso}}, \bibinfo {author} {\bibfnamefont {K.}~\bibnamefont {Destounis}},
  \bibinfo {author} {\bibfnamefont {F.}~\bibnamefont {Duque}}, \bibinfo
  {author} {\bibfnamefont {R.~P.}\ \bibnamefont {Macedo}}, \ and\ \bibinfo
  {author} {\bibfnamefont {A.}~\bibnamefont {Maselli}},\ }\href {\doibase
  10.1103/PhysRevD.105.L061501} {\bibfield  {journal} {\bibinfo  {journal}
  {Phys. Rev. D}\ }\textbf {\bibinfo {volume} {105}},\ \bibinfo {pages}
  {L061501} (\bibinfo {year} {2022})},\ \Eprint
  {http://arxiv.org/abs/2109.00005} {arXiv:2109.00005 [gr-qc]} \BibitemShut
  {NoStop}%
\bibitem [{\citenamefont {Jusufi}(2023)}]{Jusufi:2022jxu}%
  \BibitemOpen
  \bibfield  {author} {\bibinfo {author} {\bibfnamefont {K.}~\bibnamefont
  {Jusufi}},\ }\href {\doibase 10.1140/epjc/s10052-023-11264-w} {\bibfield
  {journal} {\bibinfo  {journal} {Eur. Phys. J. C}\ }\textbf {\bibinfo {volume}
  {83}},\ \bibinfo {pages} {103} (\bibinfo {year} {2023})},\ \Eprint
  {http://arxiv.org/abs/2202.00010} {arXiv:2202.00010 [gr-qc]} \BibitemShut
  {NoStop}%
\bibitem [{\citenamefont {Gondolo}\ and\ \citenamefont
  {Silk}(1999)}]{Gondolo:1999ef}%
  \BibitemOpen
  \bibfield  {author} {\bibinfo {author} {\bibfnamefont {P.}~\bibnamefont
  {Gondolo}}\ and\ \bibinfo {author} {\bibfnamefont {J.}~\bibnamefont {Silk}},\
  }\href {\doibase 10.1103/PhysRevLett.83.1719} {\bibfield  {journal} {\bibinfo
   {journal} {Phys. Rev. Lett.}\ }\textbf {\bibinfo {volume} {83}},\ \bibinfo
  {pages} {1719} (\bibinfo {year} {1999})},\ \Eprint
  {http://arxiv.org/abs/astro-ph/9906391} {arXiv:astro-ph/9906391} \BibitemShut
  {NoStop}%
\bibitem [{\citenamefont {Sadeghian}\ \emph {et~al.}(2013)\citenamefont
  {Sadeghian}, \citenamefont {Ferrer},\ and\ \citenamefont
  {Will}}]{Sadeghian:2013laa}%
  \BibitemOpen
  \bibfield  {author} {\bibinfo {author} {\bibfnamefont {L.}~\bibnamefont
  {Sadeghian}}, \bibinfo {author} {\bibfnamefont {F.}~\bibnamefont {Ferrer}}, \
  and\ \bibinfo {author} {\bibfnamefont {C.~M.}\ \bibnamefont {Will}},\ }\href
  {\doibase 10.1103/PhysRevD.88.063522} {\bibfield  {journal} {\bibinfo
  {journal} {Phys. Rev. D}\ }\textbf {\bibinfo {volume} {88}},\ \bibinfo
  {pages} {063522} (\bibinfo {year} {2013})},\ \Eprint
  {http://arxiv.org/abs/1305.2619} {arXiv:1305.2619 [astro-ph.GA]} \BibitemShut
  {NoStop}%
\bibitem [{\citenamefont {Liu}\ \emph {et~al.}(2021)\citenamefont {Liu},
  \citenamefont {Yang}, \citenamefont {Wu}, \citenamefont {Xing}, \citenamefont
  {Xu},\ and\ \citenamefont {Long}}]{Liu:2021xfb}%
  \BibitemOpen
  \bibfield  {author} {\bibinfo {author} {\bibfnamefont {D.}~\bibnamefont
  {Liu}}, \bibinfo {author} {\bibfnamefont {Y.}~\bibnamefont {Yang}}, \bibinfo
  {author} {\bibfnamefont {S.}~\bibnamefont {Wu}}, \bibinfo {author}
  {\bibfnamefont {Y.}~\bibnamefont {Xing}}, \bibinfo {author} {\bibfnamefont
  {Z.}~\bibnamefont {Xu}}, \ and\ \bibinfo {author} {\bibfnamefont {Z.-W.}\
  \bibnamefont {Long}},\ }\href {\doibase 10.1103/PhysRevD.104.104042}
  {\bibfield  {journal} {\bibinfo  {journal} {Phys. Rev. D}\ }\textbf {\bibinfo
  {volume} {104}},\ \bibinfo {pages} {104042} (\bibinfo {year} {2021})},\
  \Eprint {http://arxiv.org/abs/2104.04332} {arXiv:2104.04332 [gr-qc]}
  \BibitemShut {NoStop}%
\bibitem [{\citenamefont {Bondi}(1952)}]{Bondi:1952ni}%
  \BibitemOpen
  \bibfield  {author} {\bibinfo {author} {\bibfnamefont {H.}~\bibnamefont
  {Bondi}},\ }\href {\doibase 10.1093/mnras/112.2.195} {\bibfield  {journal}
  {\bibinfo  {journal} {Mon. Not. Roy. Astron. Soc.}\ }\textbf {\bibinfo
  {volume} {112}},\ \bibinfo {pages} {195} (\bibinfo {year}
  {1952})}\BibitemShut {NoStop}%
\bibitem [{\citenamefont {Zhao}\ \emph {et~al.}(2024)\citenamefont {Zhao},
  \citenamefont {Sun}, \citenamefont {Cao}, \citenamefont {Lin},\ and\
  \citenamefont {Qian}}]{Zhao:2023itk}%
  \BibitemOpen
  \bibfield  {author} {\bibinfo {author} {\bibfnamefont {Y.}~\bibnamefont
  {Zhao}}, \bibinfo {author} {\bibfnamefont {B.}~\bibnamefont {Sun}}, \bibinfo
  {author} {\bibfnamefont {Z.}~\bibnamefont {Cao}}, \bibinfo {author}
  {\bibfnamefont {K.}~\bibnamefont {Lin}}, \ and\ \bibinfo {author}
  {\bibfnamefont {W.-L.}\ \bibnamefont {Qian}},\ }\href {\doibase
  10.1103/PhysRevD.109.044031} {\bibfield  {journal} {\bibinfo  {journal}
  {Phys. Rev. D}\ }\textbf {\bibinfo {volume} {109}},\ \bibinfo {pages}
  {044031} (\bibinfo {year} {2024})},\ \Eprint
  {http://arxiv.org/abs/2308.15371} {arXiv:2308.15371 [gr-qc]} \BibitemShut
  {NoStop}%
\bibitem [{\citenamefont {Begeman}\ \emph {et~al.}(1991)\citenamefont
  {Begeman}, \citenamefont {Broeils},\ and\ \citenamefont
  {Sanders}}]{Begeman:1991iy}%
  \BibitemOpen
  \bibfield  {author} {\bibinfo {author} {\bibfnamefont {K.~G.}\ \bibnamefont
  {Begeman}}, \bibinfo {author} {\bibfnamefont {A.~H.}\ \bibnamefont
  {Broeils}}, \ and\ \bibinfo {author} {\bibfnamefont {R.~H.}\ \bibnamefont
  {Sanders}},\ }\href {\doibase 10.1093/mnras/249.3.523} {\bibfield  {journal}
  {\bibinfo  {journal} {Mon. Not. Roy. Astron. Soc.}\ }\textbf {\bibinfo
  {volume} {249}},\ \bibinfo {pages} {523} (\bibinfo {year}
  {1991})}\BibitemShut {NoStop}%
\bibitem [{\citenamefont {Yang}\ \emph {et~al.}(2024)\citenamefont {Yang},
  \citenamefont {Liu}, \citenamefont {\"Ovg\"un}, \citenamefont {Lambiase},\
  and\ \citenamefont {Long}}]{Yang:2023tip}%
  \BibitemOpen
  \bibfield  {author} {\bibinfo {author} {\bibfnamefont {Y.}~\bibnamefont
  {Yang}}, \bibinfo {author} {\bibfnamefont {D.}~\bibnamefont {Liu}}, \bibinfo
  {author} {\bibfnamefont {A.}~\bibnamefont {\"Ovg\"un}}, \bibinfo {author}
  {\bibfnamefont {G.}~\bibnamefont {Lambiase}}, \ and\ \bibinfo {author}
  {\bibfnamefont {Z.-W.}\ \bibnamefont {Long}},\ }\href {\doibase
  10.1140/epjc/s10052-024-12412-6} {\bibfield  {journal} {\bibinfo  {journal}
  {Eur. Phys. J. C}\ }\textbf {\bibinfo {volume} {84}},\ \bibinfo {pages} {63}
  (\bibinfo {year} {2024})},\ \Eprint {http://arxiv.org/abs/2308.05544}
  {arXiv:2308.05544 [gr-qc]} \BibitemShut {NoStop}%
\bibitem [{\citenamefont {Einasto}(1965)}]{1965On}%
  \BibitemOpen
  \bibfield  {author} {\bibinfo {author} {\bibfnamefont {J.}~\bibnamefont
  {Einasto}},\ }\href@noop {} {\bibfield  {journal} {\bibinfo  {journal} {Trudy
  Astrofizicheskogo Instituta Alma-Ata}\ }\textbf {\bibinfo {volume} {5}},\
  \bibinfo {pages} {87} (\bibinfo {year} {1965})}\BibitemShut {NoStop}%
\bibitem [{\citenamefont {Wang}\ \emph {et~al.}(2020)\citenamefont {Wang},
  \citenamefont {Bose}, \citenamefont {Frenk}, \citenamefont {Gao},
  \citenamefont {Jenkins}, \citenamefont {Springel},\ and\ \citenamefont
  {White}}]{Wang:2019ftp}%
  \BibitemOpen
  \bibfield  {author} {\bibinfo {author} {\bibfnamefont {J.}~\bibnamefont
  {Wang}}, \bibinfo {author} {\bibfnamefont {S.}~\bibnamefont {Bose}}, \bibinfo
  {author} {\bibfnamefont {C.~S.}\ \bibnamefont {Frenk}}, \bibinfo {author}
  {\bibfnamefont {L.}~\bibnamefont {Gao}}, \bibinfo {author} {\bibfnamefont
  {A.}~\bibnamefont {Jenkins}}, \bibinfo {author} {\bibfnamefont
  {V.}~\bibnamefont {Springel}}, \ and\ \bibinfo {author} {\bibfnamefont
  {S.~D.~M.}\ \bibnamefont {White}},\ }\href {\doibase
  10.1038/s41586-020-2642-9} {\bibfield  {journal} {\bibinfo  {journal}
  {Nature}\ }\textbf {\bibinfo {volume} {585}},\ \bibinfo {pages} {39}
  (\bibinfo {year} {2020})},\ \Eprint {http://arxiv.org/abs/1911.09720}
  {arXiv:1911.09720 [astro-ph.CO]} \BibitemShut {NoStop}%
\bibitem [{\citenamefont {Zhang}\ \emph {et~al.}(2021)\citenamefont {Zhang},
  \citenamefont {Zhu},\ and\ \citenamefont {Wang}}]{Zhang:2021bdr}%
  \BibitemOpen
  \bibfield  {author} {\bibinfo {author} {\bibfnamefont {C.}~\bibnamefont
  {Zhang}}, \bibinfo {author} {\bibfnamefont {T.}~\bibnamefont {Zhu}}, \ and\
  \bibinfo {author} {\bibfnamefont {A.}~\bibnamefont {Wang}},\ }\href {\doibase
  10.1103/PhysRevD.104.124082} {\bibfield  {journal} {\bibinfo  {journal}
  {Phys. Rev. D}\ }\textbf {\bibinfo {volume} {104}},\ \bibinfo {pages}
  {124082} (\bibinfo {year} {2021})},\ \Eprint
  {http://arxiv.org/abs/2111.04966} {arXiv:2111.04966 [gr-qc]} \BibitemShut
  {NoStop}%
\bibitem [{\citenamefont {Zhao}\ \emph {et~al.}(2023)\citenamefont {Zhao},
  \citenamefont {Sun}, \citenamefont {Lin},\ and\ \citenamefont
  {Cao}}]{Zhao:2023tyo}%
  \BibitemOpen
  \bibfield  {author} {\bibinfo {author} {\bibfnamefont {Y.}~\bibnamefont
  {Zhao}}, \bibinfo {author} {\bibfnamefont {B.}~\bibnamefont {Sun}}, \bibinfo
  {author} {\bibfnamefont {K.}~\bibnamefont {Lin}}, \ and\ \bibinfo {author}
  {\bibfnamefont {Z.}~\bibnamefont {Cao}},\ }\href {\doibase
  10.1103/PhysRevD.108.024070} {\bibfield  {journal} {\bibinfo  {journal}
  {Phys. Rev. D}\ }\textbf {\bibinfo {volume} {108}},\ \bibinfo {pages}
  {024070} (\bibinfo {year} {2023})},\ \Eprint
  {http://arxiv.org/abs/2303.09215} {arXiv:2303.09215 [gr-qc]} \BibitemShut
  {NoStop}%
\bibitem [{\citenamefont {Chakraborty}\ \emph {et~al.}(2025)\citenamefont
  {Chakraborty}, \citenamefont {Comp{\`e}re},\ and\ \citenamefont
  {Machet}}]{Chakraborty:2024gcr}%
  \BibitemOpen
  \bibfield  {author} {\bibinfo {author} {\bibfnamefont {S.}~\bibnamefont
  {Chakraborty}}, \bibinfo {author} {\bibfnamefont {G.}~\bibnamefont
  {Comp{\`e}re}}, \ and\ \bibinfo {author} {\bibfnamefont {L.}~\bibnamefont
  {Machet}},\ }\href {\doibase 10.1103/4p2c-rwdh} {\bibfield  {journal}
  {\bibinfo  {journal} {Phys. Rev. D}\ }\textbf {\bibinfo {volume} {112}},\
  \bibinfo {pages} {024015} (\bibinfo {year} {2025})},\ \Eprint
  {http://arxiv.org/abs/2412.14831} {arXiv:2412.14831 [gr-qc]} \BibitemShut
  {NoStop}%
\bibitem [{\citenamefont {Toshmatov}\ \emph {et~al.}(2025)\citenamefont
  {Toshmatov}, \citenamefont {Ahmedov}, \citenamefont {Boydedayev},\ and\
  \citenamefont {Ahmedov}}]{Toshmatov:2025rln}%
  \BibitemOpen
  \bibfield  {author} {\bibinfo {author} {\bibfnamefont {B.}~\bibnamefont
  {Toshmatov}}, \bibinfo {author} {\bibfnamefont {B.}~\bibnamefont {Ahmedov}},
  \bibinfo {author} {\bibfnamefont {A.}~\bibnamefont {Boydedayev}}, \ and\
  \bibinfo {author} {\bibfnamefont {B.}~\bibnamefont {Ahmedov}},\ }\href
  {\doibase 10.1103/mphy-svrk} {\bibfield  {journal} {\bibinfo  {journal}
  {Phys. Rev. D}\ }\textbf {\bibinfo {volume} {111}},\ \bibinfo {pages}
  {124058} (\bibinfo {year} {2025})}\BibitemShut {NoStop}%
\bibitem [{\citenamefont {Liu}\ \emph {et~al.}(2024)\citenamefont {Liu},
  \citenamefont {Yang},\ and\ \citenamefont {Long}}]{Liu:2024xcd}%
  \BibitemOpen
  \bibfield  {author} {\bibinfo {author} {\bibfnamefont {D.}~\bibnamefont
  {Liu}}, \bibinfo {author} {\bibfnamefont {Y.}~\bibnamefont {Yang}}, \ and\
  \bibinfo {author} {\bibfnamefont {Z.-W.}\ \bibnamefont {Long}},\ }\href
  {\doibase 10.1140/epjc/s10052-024-13096-8} {\bibfield  {journal} {\bibinfo
  {journal} {Eur. Phys. J. C}\ }\textbf {\bibinfo {volume} {84}},\ \bibinfo
  {pages} {731} (\bibinfo {year} {2024})},\ \Eprint
  {http://arxiv.org/abs/2401.09182} {arXiv:2401.09182 [gr-qc]} \BibitemShut
  {NoStop}%
\bibitem [{\citenamefont {Schwarzschild}(1916)}]{Schwarzschild:1916uq}%
  \BibitemOpen
  \bibfield  {author} {\bibinfo {author} {\bibfnamefont {K.}~\bibnamefont
  {Schwarzschild}},\ }\href@noop {} {\bibfield  {journal} {\bibinfo  {journal}
  {Sitzungsber. Preuss. Akad. Wiss. Berlin (Math. Phys. )}\ }\textbf {\bibinfo
  {volume} {1916}},\ \bibinfo {pages} {189} (\bibinfo {year} {1916})},\ \Eprint
  {http://arxiv.org/abs/physics/9905030} {arXiv:physics/9905030} \BibitemShut
  {NoStop}%
\bibitem [{\citenamefont {Daghigh}\ and\ \citenamefont
  {Kunstatter}(2024)}]{Daghigh:2023ixh}%
  \BibitemOpen
  \bibfield  {author} {\bibinfo {author} {\bibfnamefont {R.~G.}\ \bibnamefont
  {Daghigh}}\ and\ \bibinfo {author} {\bibfnamefont {G.}~\bibnamefont
  {Kunstatter}},\ }\href {\doibase 10.1103/PhysRevD.109.083004} {\bibfield
  {journal} {\bibinfo  {journal} {Phys. Rev. D}\ }\textbf {\bibinfo {volume}
  {109}},\ \bibinfo {pages} {083004} (\bibinfo {year} {2024})},\ \Eprint
  {http://arxiv.org/abs/2308.15682} {arXiv:2308.15682 [gr-qc]} \BibitemShut
  {NoStop}%
\bibitem [{\citenamefont {Daghigh}\ and\ \citenamefont
  {Kunstatter}(2022)}]{Daghigh:2022pcr}%
  \BibitemOpen
  \bibfield  {author} {\bibinfo {author} {\bibfnamefont {R.~G.}\ \bibnamefont
  {Daghigh}}\ and\ \bibinfo {author} {\bibfnamefont {G.}~\bibnamefont
  {Kunstatter}},\ }\href {\doibase 10.3847/1538-4357/ac940b} {\bibfield
  {journal} {\bibinfo  {journal} {Astrophys. J.}\ }\textbf {\bibinfo {volume}
  {940}},\ \bibinfo {pages} {33} (\bibinfo {year} {2022})},\ \Eprint
  {http://arxiv.org/abs/2206.04195} {arXiv:2206.04195 [astro-ph.GA]}
  \BibitemShut {NoStop}%
\bibitem [{\citenamefont {Tang}\ and\ \citenamefont
  {Wang}(2021)}]{Tang:2020jhx}%
  \BibitemOpen
  \bibfield  {author} {\bibinfo {author} {\bibfnamefont {M.}~\bibnamefont
  {Tang}}\ and\ \bibinfo {author} {\bibfnamefont {J.}~\bibnamefont {Wang}},\
  }\href {\doibase 10.1088/1674-1137/abc680} {\bibfield  {journal} {\bibinfo
  {journal} {Chin. Phys. C}\ }\textbf {\bibinfo {volume} {45}},\ \bibinfo
  {pages} {015110} (\bibinfo {year} {2021})},\ \Eprint
  {http://arxiv.org/abs/2005.11933} {arXiv:2005.11933 [gr-qc]} \BibitemShut
  {NoStop}%
\bibitem [{\citenamefont {Jusufi}\ \emph {et~al.}(2019)\citenamefont {Jusufi},
  \citenamefont {Jamil}, \citenamefont {Salucci}, \citenamefont {Zhu},\ and\
  \citenamefont {Haroon}}]{Jusufi:2019nrn}%
  \BibitemOpen
  \bibfield  {author} {\bibinfo {author} {\bibfnamefont {K.}~\bibnamefont
  {Jusufi}}, \bibinfo {author} {\bibfnamefont {M.}~\bibnamefont {Jamil}},
  \bibinfo {author} {\bibfnamefont {P.}~\bibnamefont {Salucci}}, \bibinfo
  {author} {\bibfnamefont {T.}~\bibnamefont {Zhu}}, \ and\ \bibinfo {author}
  {\bibfnamefont {S.}~\bibnamefont {Haroon}},\ }\href {\doibase
  10.1103/PhysRevD.100.044012} {\bibfield  {journal} {\bibinfo  {journal}
  {Phys. Rev. D}\ }\textbf {\bibinfo {volume} {100}},\ \bibinfo {pages}
  {044012} (\bibinfo {year} {2019})},\ \Eprint
  {http://arxiv.org/abs/1905.11803} {arXiv:1905.11803 [physics.gen-ph]}
  \BibitemShut {NoStop}%
\bibitem [{\citenamefont {McMillan}(2016)}]{McMillan:2016jtx}%
  \BibitemOpen
  \bibfield  {author} {\bibinfo {author} {\bibfnamefont {P.~J.}\ \bibnamefont
  {McMillan}},\ }\href {\doibase 10.1093/mnras/stw2759} {\bibfield  {journal}
  {\bibinfo  {journal} {Mon. Not. Roy. Astron. Soc.}\ }\textbf {\bibinfo
  {volume} {465}},\ \bibinfo {pages} {76} (\bibinfo {year} {2016})},\ \Eprint
  {http://arxiv.org/abs/1608.00971} {arXiv:1608.00971 [astro-ph.GA]}
  \BibitemShut {NoStop}%
\bibitem [{\citenamefont {Shen}\ \emph {et~al.}(2024)\citenamefont {Shen},
  \citenamefont {Yuan}, \citenamefont {Jiang}, \citenamefont {Tsai},
  \citenamefont {Yuan},\ and\ \citenamefont {Fan}}]{Shen:2023kkm}%
  \BibitemOpen
  \bibfield  {author} {\bibinfo {author} {\bibfnamefont {Z.-Q.}\ \bibnamefont
  {Shen}}, \bibinfo {author} {\bibfnamefont {G.-W.}\ \bibnamefont {Yuan}},
  \bibinfo {author} {\bibfnamefont {C.-Z.}\ \bibnamefont {Jiang}}, \bibinfo
  {author} {\bibfnamefont {Y.-L.~S.}\ \bibnamefont {Tsai}}, \bibinfo {author}
  {\bibfnamefont {Q.}~\bibnamefont {Yuan}}, \ and\ \bibinfo {author}
  {\bibfnamefont {Y.-Z.}\ \bibnamefont {Fan}},\ }\href {\doibase
  10.1093/mnras/stad3282} {\bibfield  {journal} {\bibinfo  {journal} {Mon. Not.
  Roy. Astron. Soc.}\ }\textbf {\bibinfo {volume} {527}},\ \bibinfo {pages}
  {3196} (\bibinfo {year} {2024})},\ \Eprint {http://arxiv.org/abs/2303.09284}
  {arXiv:2303.09284 [astro-ph.GA]} \BibitemShut {NoStop}%
\bibitem [{\citenamefont {Nampalliwar}\ \emph {et~al.}(2021)\citenamefont
  {Nampalliwar}, \citenamefont {Kumar}, \citenamefont {Jusufi}, \citenamefont
  {Wu}, \citenamefont {Jamil},\ and\ \citenamefont
  {Salucci}}]{Nampalliwar:2021tyz}%
  \BibitemOpen
  \bibfield  {author} {\bibinfo {author} {\bibfnamefont {S.}~\bibnamefont
  {Nampalliwar}}, \bibinfo {author} {\bibfnamefont {S.}~\bibnamefont {Kumar}},
  \bibinfo {author} {\bibfnamefont {K.}~\bibnamefont {Jusufi}}, \bibinfo
  {author} {\bibfnamefont {Q.}~\bibnamefont {Wu}}, \bibinfo {author}
  {\bibfnamefont {M.}~\bibnamefont {Jamil}}, \ and\ \bibinfo {author}
  {\bibfnamefont {P.}~\bibnamefont {Salucci}},\ }\href {\doibase
  10.3847/1538-4357/ac05cc} {\bibfield  {journal} {\bibinfo  {journal}
  {Astrophys. J.}\ }\textbf {\bibinfo {volume} {916}},\ \bibinfo {pages} {116}
  (\bibinfo {year} {2021})},\ \Eprint {http://arxiv.org/abs/2103.12439}
  {arXiv:2103.12439 [astro-ph.HE]} \BibitemShut {NoStop}%
\bibitem [{\citenamefont {Davis}\ \emph {et~al.}(1972)\citenamefont {Davis},
  \citenamefont {Ruffini},\ and\ \citenamefont {Tiomno}}]{Davis:1972ud}%
  \BibitemOpen
  \bibfield  {author} {\bibinfo {author} {\bibfnamefont {M.}~\bibnamefont
  {Davis}}, \bibinfo {author} {\bibfnamefont {R.}~\bibnamefont {Ruffini}}, \
  and\ \bibinfo {author} {\bibfnamefont {J.}~\bibnamefont {Tiomno}},\ }\href
  {\doibase 10.1103/PhysRevD.5.2932} {\bibfield  {journal} {\bibinfo  {journal}
  {Phys. Rev. D}\ }\textbf {\bibinfo {volume} {5}},\ \bibinfo {pages} {2932}
  (\bibinfo {year} {1972})}\BibitemShut {NoStop}%
\bibitem [{\citenamefont {{Futterman}}\ \emph {et~al.}(1988)\citenamefont
  {{Futterman}}, \citenamefont {{Handler}},\ and\ \citenamefont
  {{Matzner}}}]{1988sfbh.book.....F}%
  \BibitemOpen
  \bibfield  {author} {\bibinfo {author} {\bibfnamefont {J.~A.~H.}\
  \bibnamefont {{Futterman}}}, \bibinfo {author} {\bibfnamefont {F.~A.}\
  \bibnamefont {{Handler}}}, \ and\ \bibinfo {author} {\bibfnamefont {R.~A.}\
  \bibnamefont {{Matzner}}},\ }\href@noop {} {\emph {\bibinfo {title}
  {{Scattering from black holes}}}}\ (\bibinfo  {publisher} {{Cambridge
  University Press}},\ \bibinfo {year} {1988})\BibitemShut {NoStop}%
\bibitem [{\citenamefont {Regge}\ and\ \citenamefont
  {Wheeler}(1957)}]{Regge:1957td}%
  \BibitemOpen
  \bibfield  {author} {\bibinfo {author} {\bibfnamefont {T.}~\bibnamefont
  {Regge}}\ and\ \bibinfo {author} {\bibfnamefont {J.~A.}\ \bibnamefont
  {Wheeler}},\ }\href {\doibase 10.1103/PhysRev.108.1063} {\bibfield  {journal}
  {\bibinfo  {journal} {Phys. Rev.}\ }\textbf {\bibinfo {volume} {108}},\
  \bibinfo {pages} {1063} (\bibinfo {year} {1957})}\BibitemShut {NoStop}%
\bibitem [{\citenamefont {Moderski}\ and\ \citenamefont
  {Rogatko}(2001{\natexlab{a}})}]{PhysRevD.63.084014}%
  \BibitemOpen
  \bibfield  {author} {\bibinfo {author} {\bibfnamefont {R.}~\bibnamefont
  {Moderski}}\ and\ \bibinfo {author} {\bibfnamefont {M.}~\bibnamefont
  {Rogatko}},\ }\href {\doibase 10.1103/PhysRevD.63.084014} {\bibfield
  {journal} {\bibinfo  {journal} {Phys. Rev. D}\ }\textbf {\bibinfo {volume}
  {63}},\ \bibinfo {pages} {084014} (\bibinfo {year}
  {2001}{\natexlab{a}})}\BibitemShut {NoStop}%
\bibitem [{\citenamefont {Moderski}\ and\ \citenamefont
  {Rogatko}(2001{\natexlab{b}})}]{PhysRevD.64.044024}%
  \BibitemOpen
  \bibfield  {author} {\bibinfo {author} {\bibfnamefont {R.}~\bibnamefont
  {Moderski}}\ and\ \bibinfo {author} {\bibfnamefont {M.}~\bibnamefont
  {Rogatko}},\ }\href {\doibase 10.1103/PhysRevD.64.044024} {\bibfield
  {journal} {\bibinfo  {journal} {Phys. Rev. D}\ }\textbf {\bibinfo {volume}
  {64}},\ \bibinfo {pages} {044024} (\bibinfo {year}
  {2001}{\natexlab{b}})}\BibitemShut {NoStop}%
\bibitem [{\citenamefont {Moderski}\ and\ \citenamefont
  {Rogatko}(2005)}]{PhysRevD.72.044027}%
  \BibitemOpen
  \bibfield  {author} {\bibinfo {author} {\bibfnamefont {R.}~\bibnamefont
  {Moderski}}\ and\ \bibinfo {author} {\bibfnamefont {M.}~\bibnamefont
  {Rogatko}},\ }\href {\doibase 10.1103/PhysRevD.72.044027} {\bibfield
  {journal} {\bibinfo  {journal} {Phys. Rev. D}\ }\textbf {\bibinfo {volume}
  {72}},\ \bibinfo {pages} {044027} (\bibinfo {year} {2005})}\BibitemShut
  {NoStop}%
\bibitem [{\citenamefont {Konoplya}\ and\ \citenamefont
  {Zhidenko}(2011{\natexlab{b}})}]{RevModPhys.83.793}%
  \BibitemOpen
  \bibfield  {author} {\bibinfo {author} {\bibfnamefont {R.~A.}\ \bibnamefont
  {Konoplya}}\ and\ \bibinfo {author} {\bibfnamefont {A.}~\bibnamefont
  {Zhidenko}},\ }\href {\doibase 10.1103/RevModPhys.83.793} {\bibfield
  {journal} {\bibinfo  {journal} {Rev. Mod. Phys.}\ }\textbf {\bibinfo {volume}
  {83}},\ \bibinfo {pages} {793} (\bibinfo {year}
  {2011}{\natexlab{b}})}\BibitemShut {NoStop}%
\bibitem [{\citenamefont {Berti}\ \emph {et~al.}(2007)\citenamefont {Berti},
  \citenamefont {Cardoso}, \citenamefont {Gonz\'alez},\ and\ \citenamefont
  {Sperhake}}]{PhysRevD.75.124017}%
  \BibitemOpen
  \bibfield  {author} {\bibinfo {author} {\bibfnamefont {E.}~\bibnamefont
  {Berti}}, \bibinfo {author} {\bibfnamefont {V.}~\bibnamefont {Cardoso}},
  \bibinfo {author} {\bibfnamefont {J.~A.}\ \bibnamefont {Gonz\'alez}}, \ and\
  \bibinfo {author} {\bibfnamefont {U.}~\bibnamefont {Sperhake}},\ }\href
  {\doibase 10.1103/PhysRevD.75.124017} {\bibfield  {journal} {\bibinfo
  {journal} {Phys. Rev. D}\ }\textbf {\bibinfo {volume} {75}},\ \bibinfo
  {pages} {124017} (\bibinfo {year} {2007})}\BibitemShut {NoStop}%
\bibitem [{\citenamefont {Chowdhury}\ and\ \citenamefont
  {Banerjee}(2020)}]{Chowdhury:2020rfj}%
  \BibitemOpen
  \bibfield  {author} {\bibinfo {author} {\bibfnamefont {A.}~\bibnamefont
  {Chowdhury}}\ and\ \bibinfo {author} {\bibfnamefont {N.}~\bibnamefont
  {Banerjee}},\ }\href {\doibase 10.1103/PhysRevD.102.124051} {\bibfield
  {journal} {\bibinfo  {journal} {Phys. Rev. D}\ }\textbf {\bibinfo {volume}
  {102}},\ \bibinfo {pages} {124051} (\bibinfo {year} {2020})},\ \Eprint
  {http://arxiv.org/abs/2006.16522} {arXiv:2006.16522 [gr-qc]} \BibitemShut
  {NoStop}%
\bibitem [{\citenamefont {Leaver}(1985)}]{Leaver:1985ax}%
  \BibitemOpen
  \bibfield  {author} {\bibinfo {author} {\bibfnamefont {E.~W.}\ \bibnamefont
  {Leaver}},\ }\href {\doibase 10.1098/rspa.1985.0119} {\bibfield  {journal}
  {\bibinfo  {journal} {Proc. Roy. Soc. Lond. A}\ }\textbf {\bibinfo {volume}
  {402}},\ \bibinfo {pages} {285} (\bibinfo {year} {1985})}\BibitemShut
  {NoStop}%
\bibitem [{\citenamefont {Leaver}(1990)}]{Leaver:1990zz}%
  \BibitemOpen
  \bibfield  {author} {\bibinfo {author} {\bibfnamefont {E.~W.}\ \bibnamefont
  {Leaver}},\ }\href {\doibase 10.1103/PhysRevD.41.2986} {\bibfield  {journal}
  {\bibinfo  {journal} {Phys. Rev. D}\ }\textbf {\bibinfo {volume} {41}},\
  \bibinfo {pages} {2986} (\bibinfo {year} {1990})}\BibitemShut {NoStop}%
\bibitem [{\citenamefont {Berti}\ \emph {et~al.}(2009)\citenamefont {Berti},
  \citenamefont {Cardoso},\ and\ \citenamefont {Starinets}}]{Berti:2009kk}%
  \BibitemOpen
  \bibfield  {author} {\bibinfo {author} {\bibfnamefont {E.}~\bibnamefont
  {Berti}}, \bibinfo {author} {\bibfnamefont {V.}~\bibnamefont {Cardoso}}, \
  and\ \bibinfo {author} {\bibfnamefont {A.~O.}\ \bibnamefont {Starinets}},\
  }\href {\doibase 10.1088/0264-9381/26/16/163001} {\bibfield  {journal}
  {\bibinfo  {journal} {Class. Quant. Grav.}\ }\textbf {\bibinfo {volume}
  {26}},\ \bibinfo {pages} {163001} (\bibinfo {year} {2009})},\ \Eprint
  {http://arxiv.org/abs/0905.2975} {arXiv:0905.2975 [gr-qc]} \BibitemShut
  {NoStop}%
\bibitem [{\citenamefont {Anacleto}\ \emph {et~al.}(2021)\citenamefont
  {Anacleto}, \citenamefont {Campos}, \citenamefont {Brito},\ and\
  \citenamefont {Passos}}]{Anacleto:2021qoe}%
  \BibitemOpen
  \bibfield  {author} {\bibinfo {author} {\bibfnamefont {M.~A.}\ \bibnamefont
  {Anacleto}}, \bibinfo {author} {\bibfnamefont {J.~A.~V.}\ \bibnamefont
  {Campos}}, \bibinfo {author} {\bibfnamefont {F.~A.}\ \bibnamefont {Brito}}, \
  and\ \bibinfo {author} {\bibfnamefont {E.}~\bibnamefont {Passos}},\ }\href
  {\doibase 10.1016/j.aop.2021.168662} {\bibfield  {journal} {\bibinfo
  {journal} {Annals Phys.}\ }\textbf {\bibinfo {volume} {434}},\ \bibinfo
  {pages} {168662} (\bibinfo {year} {2021})},\ \Eprint
  {http://arxiv.org/abs/2108.04998} {arXiv:2108.04998 [gr-qc]} \BibitemShut
  {NoStop}%
\bibitem [{\citenamefont {Siqueira}\ and\ \citenamefont
  {Richartz}(2022)}]{Siqueira:2022tbc}%
  \BibitemOpen
  \bibfield  {author} {\bibinfo {author} {\bibfnamefont {P.~H.~C.}\
  \bibnamefont {Siqueira}}\ and\ \bibinfo {author} {\bibfnamefont
  {M.}~\bibnamefont {Richartz}},\ }\href {\doibase 10.1103/PhysRevD.106.024046}
  {\bibfield  {journal} {\bibinfo  {journal} {Phys. Rev. D}\ }\textbf {\bibinfo
  {volume} {106}},\ \bibinfo {pages} {024046} (\bibinfo {year} {2022})},\
  \Eprint {http://arxiv.org/abs/2205.00556} {arXiv:2205.00556 [gr-qc]}
  \BibitemShut {NoStop}%
\bibitem [{\citenamefont {Berti}\ \emph
  {et~al.}(2006{\natexlab{a}})\citenamefont {Berti}, \citenamefont {Cardoso},\
  and\ \citenamefont {Casals}}]{Berti:2005gp}%
  \BibitemOpen
  \bibfield  {author} {\bibinfo {author} {\bibfnamefont {E.}~\bibnamefont
  {Berti}}, \bibinfo {author} {\bibfnamefont {V.}~\bibnamefont {Cardoso}}, \
  and\ \bibinfo {author} {\bibfnamefont {M.}~\bibnamefont {Casals}},\ }\href
  {\doibase 10.1103/PhysRevD.73.109902} {\bibfield  {journal} {\bibinfo
  {journal} {Phys. Rev. D}\ }\textbf {\bibinfo {volume} {73}},\ \bibinfo
  {pages} {024013} (\bibinfo {year} {2006}{\natexlab{a}})},\ \bibinfo {note}
  {[Erratum: Phys.Rev.D 73, 109902 (2006)]},\ \Eprint
  {http://arxiv.org/abs/gr-qc/0511111} {arXiv:gr-qc/0511111} \BibitemShut
  {NoStop}%
\bibitem [{\citenamefont {Yoshino}\ and\ \citenamefont
  {Kodama}(2014)}]{Yoshino:2013ofa}%
  \BibitemOpen
  \bibfield  {author} {\bibinfo {author} {\bibfnamefont {H.}~\bibnamefont
  {Yoshino}}\ and\ \bibinfo {author} {\bibfnamefont {H.}~\bibnamefont
  {Kodama}},\ }\href {\doibase 10.1093/ptep/ptu029} {\bibfield  {journal}
  {\bibinfo  {journal} {PTEP}\ }\textbf {\bibinfo {volume} {2014}},\ \bibinfo
  {pages} {043E02} (\bibinfo {year} {2014})},\ \Eprint
  {http://arxiv.org/abs/1312.2326} {arXiv:1312.2326 [gr-qc]} \BibitemShut
  {NoStop}%
\bibitem [{\citenamefont {Nollert}(1993)}]{Nollert:1993zz}%
  \BibitemOpen
  \bibfield  {author} {\bibinfo {author} {\bibfnamefont {H.-P.}\ \bibnamefont
  {Nollert}},\ }\href {\doibase 10.1103/PhysRevD.47.5253} {\bibfield  {journal}
  {\bibinfo  {journal} {Phys. Rev. D}\ }\textbf {\bibinfo {volume} {47}},\
  \bibinfo {pages} {5253} (\bibinfo {year} {1993})}\BibitemShut {NoStop}%
\bibitem [{\citenamefont {Zhidenko}(2006)}]{Zhidenko:2006rs}%
  \BibitemOpen
  \bibfield  {author} {\bibinfo {author} {\bibfnamefont {A.}~\bibnamefont
  {Zhidenko}},\ }\href {\doibase 10.1103/PhysRevD.74.064017} {\bibfield
  {journal} {\bibinfo  {journal} {Phys. Rev. D}\ }\textbf {\bibinfo {volume}
  {74}},\ \bibinfo {pages} {064017} (\bibinfo {year} {2006})},\ \Eprint
  {http://arxiv.org/abs/gr-qc/0607133} {arXiv:gr-qc/0607133} \BibitemShut
  {NoStop}%
\bibitem [{\citenamefont {Iyer}(1987)}]{Iyer:1986nq}%
  \BibitemOpen
  \bibfield  {author} {\bibinfo {author} {\bibfnamefont {S.}~\bibnamefont
  {Iyer}},\ }\href {\doibase 10.1103/PhysRevD.35.3632} {\bibfield  {journal}
  {\bibinfo  {journal} {Phys. Rev. D}\ }\textbf {\bibinfo {volume} {35}},\
  \bibinfo {pages} {3632} (\bibinfo {year} {1987})}\BibitemShut {NoStop}%
\bibitem [{\citenamefont {Berti}\ \emph
  {et~al.}(2006{\natexlab{b}})\citenamefont {Berti}, \citenamefont {Cardoso},\
  and\ \citenamefont {Will}}]{Berti:2005ys}%
  \BibitemOpen
  \bibfield  {author} {\bibinfo {author} {\bibfnamefont {E.}~\bibnamefont
  {Berti}}, \bibinfo {author} {\bibfnamefont {V.}~\bibnamefont {Cardoso}}, \
  and\ \bibinfo {author} {\bibfnamefont {C.~M.}\ \bibnamefont {Will}},\ }\href
  {\doibase 10.1103/PhysRevD.73.064030} {\bibfield  {journal} {\bibinfo
  {journal} {Phys. Rev. D}\ }\textbf {\bibinfo {volume} {73}},\ \bibinfo
  {pages} {064030} (\bibinfo {year} {2006}{\natexlab{b}})},\ \Eprint
  {http://arxiv.org/abs/gr-qc/0512160} {arXiv:gr-qc/0512160} \BibitemShut
  {NoStop}%
\bibitem [{\citenamefont {Zhang}\ \emph {et~al.}(2022)\citenamefont {Zhang},
  \citenamefont {Zhu}, \citenamefont {Fang},\ and\ \citenamefont
  {Wang}}]{Zhang:2022roh}%
  \BibitemOpen
  \bibfield  {author} {\bibinfo {author} {\bibfnamefont {C.}~\bibnamefont
  {Zhang}}, \bibinfo {author} {\bibfnamefont {T.}~\bibnamefont {Zhu}}, \bibinfo
  {author} {\bibfnamefont {X.}~\bibnamefont {Fang}}, \ and\ \bibinfo {author}
  {\bibfnamefont {A.}~\bibnamefont {Wang}},\ }\href {\doibase
  10.1016/j.dark.2022.101078} {\bibfield  {journal} {\bibinfo  {journal} {Phys.
  Dark Univ.}\ }\textbf {\bibinfo {volume} {37}},\ \bibinfo {pages} {101078}
  (\bibinfo {year} {2022})},\ \Eprint {http://arxiv.org/abs/2201.11352}
  {arXiv:2201.11352 [gr-qc]} \BibitemShut {NoStop}%
\bibitem [{\citenamefont {Shi}\ \emph {et~al.}(2019)\citenamefont {Shi},
  \citenamefont {Bao}, \citenamefont {Wang}, \citenamefont {Zhang},
  \citenamefont {Hu}, \citenamefont {Sesana}, \citenamefont {Barausse},
  \citenamefont {Mei},\ and\ \citenamefont {Luo}}]{Shi:2019hqa}%
  \BibitemOpen
  \bibfield  {author} {\bibinfo {author} {\bibfnamefont {C.}~\bibnamefont
  {Shi}}, \bibinfo {author} {\bibfnamefont {J.}~\bibnamefont {Bao}}, \bibinfo
  {author} {\bibfnamefont {H.}~\bibnamefont {Wang}}, \bibinfo {author}
  {\bibfnamefont {J.-d.}\ \bibnamefont {Zhang}}, \bibinfo {author}
  {\bibfnamefont {Y.}~\bibnamefont {Hu}}, \bibinfo {author} {\bibfnamefont
  {A.}~\bibnamefont {Sesana}}, \bibinfo {author} {\bibfnamefont
  {E.}~\bibnamefont {Barausse}}, \bibinfo {author} {\bibfnamefont
  {J.}~\bibnamefont {Mei}}, \ and\ \bibinfo {author} {\bibfnamefont
  {J.}~\bibnamefont {Luo}},\ }\href {\doibase 10.1103/PhysRevD.100.044036}
  {\bibfield  {journal} {\bibinfo  {journal} {Phys. Rev. D}\ }\textbf {\bibinfo
  {volume} {100}},\ \bibinfo {pages} {044036} (\bibinfo {year} {2019})},\
  \Eprint {http://arxiv.org/abs/1902.08922} {arXiv:1902.08922 [gr-qc]}
  \BibitemShut {NoStop}%
\end{thebibliography}%
\end{document}